\documentclass[aps, preprint, nofootinbib,preprintnumbers,eqsecnum,superscriptaddress,sort]{revtex4}
\pdfoutput=1
\usepackage{amssymb,amsmath,amsfonts,amsthm,latexsym,mathrsfs}
\usepackage{graphics, graphicx,color,epsfig,subfigure}
\usepackage[
      colorlinks=true,
      linkcolor=blue,
      urlcolor=blue,
      filecolor=black,
      citecolor=red,
      pdfstartview=FitV,
      pdftitle={},
        pdfauthor={Gary Horowitz, Jorge Santos},
        pdfsubject={},
        pdfkeywords={},
        pdfpagemode=None,
        bookmarksopen=true
      ]{hyperref}
\numberwithin{equation}{section}
\def \be {\begin{equation}}
\def \ee {\end{equation}}
\def \bea {\begin{eqnarray}}
\def \eea {\end{eqnarray}}
\def \dd {\mathrm{d}}

\newcommand{\bln}{\begin{align}}
\newcommand{\eln}{\end{align}}
\newcommand{\bst}{\begin{split}}
\newcommand{\est}{\end{split}}
\newcommand{\bi}{\begin{itemize}}
\newcommand{\ei}{\end{itemize}}
\newcommand{\ben}{\begin{enumerate}}
\newcommand{\een}{\end{enumerate}}

\def\eeq{\end{equation}}

\newcommand\ergo{\rm ergo}

\newcommand\smallO{
  \mathchoice
    {{\scriptstyle\mathcal{O}}}
    {{\scriptstyle\mathcal{O}}}
    {{\scriptscriptstyle\mathcal{O}}}
    {\scalebox{.7}{$\scriptscriptstyle\mathcal{O}$}}
  }

\begin{document}

\title{Attempts at vacuum counterexamples to \\
cosmic censorship in AdS}

\author{Toby~Crisford}
\email{tc393@cam.ac.uk}
\affiliation{Department of Applied Mathematics and Theoretical Physics, University of Cambridge, Wilberforce Road, Cambridge CB3 0WA, UK}

\author{Gary~T.~Horowitz}
\email{horowitz@ucsb.edu}
\affiliation{Department of Physics, UCSB, Santa Barbara, CA 93106}

\author{Jorge~E.~Santos}
\email{jss55@cam.ac.uk}
\affiliation{Department of Applied Mathematics and Theoretical Physics, University of Cambridge, Wilberforce Road, Cambridge CB3 0WA, UK \vspace{1 cm}}

\begin{abstract}\noindent{

We consider vacuum solutions of four dimensional general relativity with $\Lambda < 0$. We numerically construct stationary solutions that asymptotically approach a boundary metric with differential rotation. Smooth solutions only exist up to a critical  rotation. We thus argue that increasing the differential rotation by a finite amount will cause the curvature to grow without bound.   This holds for both zero and nonzero temperature, and both compact and noncompact boundaries.   However,  the boundary metric always develops an ergoregion before reaching the critical rotation, which probably means that the energy is unbounded from below for these counterexamples to cosmic censorship.
}

\end{abstract}
\maketitle

\tableofcontents

\section{Introduction and Summary}

Cosmic censorship \cite{Penrose:1969pc} remains one of the most important open problems in classical general relativity.  We will be interested in the weak form of this conjecture which roughly says that one cannot generically form regions of arbitrarily large curvature that are visible to infinity. Although it is usually discussed in the context of four-dimensional asymptotically flat spacetimes, there has been recent interest in higher dimensions and other boundary conditions. We will stay in  four-dimensions, but consider asymptotically anti-de Sitter (AdS) spacetimes. This is motivated by gauge/gravity duality which relates gravity with this boundary condition to a nongravitational gauge theory \cite{Aharony:1999ti}.

In this context, a class of counterexamples to cosmic censorship has recently been found if one adds a Maxwell field \cite{Horowitz:2016ezu,Crisford:2017zpi}.  In 
asymptotically AdS spacetimes, one is free to specify a boundary (conformal) metric and boundary values of any matter fields that might be present. If one assumes certain  static profiles for the asymptotic vector potential and multiplies them by an overall amplitude $a$, it turns out that static, nonsingular, $T=0$ solutions only exist up to a maximum amplitude $a_{\max}$ (which depends on the profile). These are static self-gravitating electric fields which become singular as $ a\to a_{\max}$. It was then shown \cite{Crisford:2017zpi} that if one considers a time dependent boundary condition where the amplitude  starts with $a<a_{\max}$
   and ends with $a>a_{\max}$, the electric field and hence the curvature grow as a power law in time over a large region visible from infinity. 
  Although the singularity does not form in finite time, these examples clearly violate the spirit of cosmic censorship.

In this paper, we attempt to construct vacuum  analogs of these counterexamples. Instead of adding a Maxwell field, we will add differential rotation to the boundary metric and construct smooth stationary solutions which approach this asymptotic geometry.
 A similar setup was studied in \cite{Markeviciute:2017jcp}, where a dipolar differential rotation was added at the conformal boundary using global coordinates. We will show that if one keeps the profile of the differential rotation fixed, but increases the overall amplitude, smooth solutions only exist up to a finite maximum amplitude $a_{\max}$. As before, we expect that in the time dependent case where the amplitude is increased from $a<a_{\max}$ to $a>a_{\max}$, the curvature will again increase without bound.

We consider both $T = 0$ and $T>0$ solutions, and boundaries that are both compact and noncompact. In all  cases the results are qualitatively the same. There is a finite amplitude, $a_{\max}$, beyond which smooth stationary solutions do not exist. However, before reaching $a_{\max}$, both the boundary metric and bulk spacetime develop an ergoregion, \emph{i.e.}, a region of spacetime where the time translation Killing vector becomes spacelike\footnote{Note that the existence of an ergoregion  is a conformally invariant property of the boundary metric.}.  If $a_{\ergo}$ denotes the amplitude at which the ergoregion first forms, we will see that  $a_{\max} -a_{\ergo}$  can be made as small as one likes by varying the profile or temperature, but it is always positive. The existence of an ergoregion causes two problems  which we now discuss.  

First, spacetimes with ergoregions in AdS may be unstable due to superradiant scattering.  This is known to happen when the ergoregion surrounds a spherical black hole.  Certain modes can scatter off the black hole and return with greater amplitude. They then reflect off infinity and scatter off the black hole repeatedly, leading to an instability.  The endpoint of this instability is not known although there has been some remarkable recent numerical progress  \cite{Chesler:2018txn}.  It may in fact violate weak cosmic censorship in vacuum \cite{Dias:2015rxy,Niehoff:2015oga}. In our case,  the ergoregion is in the asymptotic region and it is not clear if a superradiant instability exists, since ingoing waves are partially absorbed by the horizon and return with smaller amplitude. This can compensate for the enhanced scattering off the ergoregion. However, given the results in \cite{Green:2015kur}, it is likely that our solutions are also unstable.


A more serious problem is that the energy is likely to be unbounded from below.  The existing proofs of positive energy in AdS \cite{Gibbons:1982jg,Cheng:2005wk,Xie:2007qp,Chrusciel:2018lpj} do not apply to boundary metrics with ergoregions. We will discuss this in section V and give arguments  that the energy is probably not bounded from below. Thus these vacuum counterexamples to cosmic censorship are less interesting than the electromagnetic counterexamples, and do not have the same status.

This is good news for the suggested connection between cosmic censorship and the weak gravity conjecture \cite{ArkaniHamed:2006dz}. For the electromagnetic counterexamples,  it was shown that  adding a charged scalar field with mass $m$ and charge $q$ causes the Einstein-Maxwell solutions that violate cosmic censorship to become  unstable  if $q/m$ is large enough \cite{Crisford:2017gsb}. Furthermore, the instability results in a nonzero scalar field and one can no longer violate cosmic censorship. Surprisingly,  the minimum value of $q/m$ to preserve cosmic censorship turns out to be precisely that predicted by the weak gravity conjecture adapted to AdS \cite{Crisford:2017gsb}. Since our current vacuum counterexamples to cosmic censorship probably have a similar effect on the geometry but do not involve any electromagnetic fields, they could not be removed by invoking the weak gravity conjecture.

Our stationary solutions with $T > 0$ have a standard black hole horizon in the interior. But the infrared behavior of the $T=0$ solutions depends on the fall-off of the differential rotation. If it falls-off faster than $1/r$, the effects of the rotation die off as one moves into the bulk and the solution has a standard Poincar\'e horizon. If it falls off like $1/r$, there is a new extremal horizon which we will describe explicitly. If the profile is exactly $1/r$, the solution has an additional scaling symmetry and can be written analytically. We call this the ``spinning top" solution since the angular momentum density is concentrated at the origin. It can be viewed as the vacuum analog of the analytic ``point charge" solution found in \cite{Horowitz:2014gva}.

In many cases, the boundary metrics and bulk black holes that we construct will be axisymmetric as well as stationary.  This raises the possibility that there may be a nonaxisymmetric, stationary black holes with $a > a_{\max}$.  This is not possible in asymptotically flat spacetimes, but might occur in AdS \cite{Dias:2015rxy}. To check this, we will study some cases where the only symmetry of the boundary metric is the stationary Killing field. We again find there is a maximum amplitude for smooth solutions.


In the next section we briefly review how to numerically construct stationary vacuum solutions. Sections III and IV give some further details on the construction and contain our main results for stationary solutions with rotating planar (III) or compact (IV)  boundary metrics. We show there is a maximum amplitude and describe some properties of the solutions. In the last section we give arguments that the energy is unbounded from below, and discuss some implications for the dual field theory.

\section{\label{sec:turck}Constructing general rotating defects}
We will be interested in finding stationary, asymptotically AdS solutions of  Einstein's equation:
\begin{equation}
R_{ab}+\frac{3}{L^2}g_{ab}=0\,,
\label{eq:einstein}
\end{equation}
where $L$ is the AdS length scale and $R_{ab}$ the four-dimensional Ricci tensor associated with the metric $g_{ab}$. Throughout this manuscript we work with $G_4=1$.

 In order to find solutions to \eqref{eq:einstein}  numerically we will use the so called DeTurck method, which was first presented in \cite{Headrick:2009pv} and reviewed in great detail in \cite{Wiseman:2011by,Dias:2015nua}. The idea is to consider the following modification of Eq~(\ref{eq:einstein})
\begin{equation}
R_{ab}+\frac{3}{L^2}g_{ab}-\nabla_{(a}\xi_{b)}=0\,,
\label{eq:einsteindeturck}
\end{equation}
where $\xi^a = [\Gamma(g)^{a}_{bc}-\Gamma(\bar{g})^{a}_{bc}]g^{bc}$ is the so called DeTurck vector, $\Gamma(\mathfrak{g})$ is the Levi-Civita connection associated to a metric $\mathfrak{g}$ and $\bar{g}$ is a reference metric which will be related to our choice of gauge. In terms of spacetime coordinates, we have $\xi^a = -\Box x^a+H^a$, where $H^a \equiv -\Gamma(\bar{g})^{a}_{bc}g^{bc}$ does not explicitly depend on derivatives of $g$. Solutions of the Einstein equation (\ref{eq:einstein}) will be solutions of the Einstein-DeTurck equation (\ref{eq:einsteindeturck}) with $\bar{g}=g$, however, the converse might not always be true. That is to say, it is not clear whether solutions of (\ref{eq:einsteindeturck}) will necessarily be solutions of (\ref{eq:einstein}), \emph{i.e.} solutions with $\xi^a\neq0$ might exist.

However, most of the boundary metrics we will consider are stationary, axially symmetric and have a $(t,\phi)\to-(t,\phi)$ reflection symmetry, all of which extend into the bulk. Under these symmetries, Figueras and Wiseman have shown in \cite{Figueras:2016nmo} that $\xi$ must vanish on solutions of (\ref{eq:einsteindeturck}). The advantage of solving Eq.~(\ref{eq:einsteindeturck}) instead of Eq.~(\ref{eq:einstein}) is immense, since the former represents a system of elliptic equations which can be readily solved using a standard relaxation procedure. The gauge, which is dynamically determined during the numerical procedure, is given by $\xi=0\Rightarrow \Box x^a = H^a$.
\section{Planar solutions}

In this section we discuss
stationary, axisymmetric solutions to \eqref{eq:einstein} with boundary metrics of the form
\begin{equation}
\dd s^2_\partial = -\dd t^2+\dd r^2+r^2 [\dd\phi - \omega(r) \dd t]^2\;,
\label{eq:bndmetric}
\end{equation}
with
\be
\omega(r) = a\,p(r).
\ee
These metrics describe geometries with differential rotation with an amplitude $a$ and profile $p(r)$. We will demand that $p(r) \to 0$ as $r\to \infty$.

\subsection{Zero temperature solutions}
We start by considering solutions at zero temperature.   It is clear from \eqref{eq:bndmetric} that $\omega(r)$ must have (mass) dimension one. So if it falls off like $a/r^n$, the dimension of $a$ must be $1-n$. Thus for $n > 1$, turning on $a$  represents an irrelevant deformation of the boundary metric and the solution should have a standard Poincar\'e horizon. We will see below that this is indeed the case. For $n < 1$, turning on $a$  is a relevant deformation and the solution will be very different in the infrared. For $n =1$, $a$ is dimensionless and corresponds to a marginal deformation. In this case, the extremal horizon is deformed in a way that we now describe.

\subsubsection{\label{sec:spinning}Holographic spinning top}

When $\omega(r) = a/r$ everywhere, the boundary metric has a new scaling symmetry:
$t \to \lambda t, \ r\to \lambda r$. In fact, the conformal  metric is invariant under an $SO(2,1)\times SO(2)$ subgroup of the full $SO(3,2)$ conformal group of flat space. The corresponding bulk solution also has this extended symmetry and can be described analytically. In fact, it can be obtained by a double Wick rotation of a hyperbolic Taub-Nut black hole in AdS$_4$ \cite{Chamblin:1998pz}, which is an algebraically special solution in the Petrov classification. 

The resulting solution can be written
\begin{equation}\label{eq:taubnut}
\mathrm{d}s^2=\frac{L^2}{\eta^2}\left[H(\eta)\left(-\rho^2\mathrm{d}t^2+\frac{\mathrm{d}\rho^2}{\rho^2}\right)+\frac{y_+^2\,H(\eta)\mathrm{d}\eta^2}{(1-\eta)G(\eta)}+\frac{G(\eta)(1-\eta)}{H(\eta)}(\mathrm{d}\phi-2n\rho\mathrm{d}t)^2\right]
\end{equation}
with
\begin{equation}
G(\eta)=n^2 \left(3 n^2-1\right) \eta^3+\left(6 n^2-1\right) \eta^2 y_+^2+\left(1+\eta+\eta^2\right) y_+^4\,,\quad\text{and}\quad H(\eta)=y_+^2+n^2\,\eta^2\,.
\end{equation}
 Here, $\eta\in(0,1]$ with $\eta = 0$ being the asymptotic boundary and $\eta =1$ is the axis of rotation for $\partial_\phi$. The first term in parenthesis on the right is AdS$_2$ with a horizon at $\rho=0$. This null surface defines a degenerate horizon for the full four dimensional spacetime, so the solution has zero temperature.  The solution depends on two parameters, $y_+$ and $n$, and for generic values of them, there is a conical singularity along the rotation axis $\eta =1$.
  This can be readily avoided by demanding
\begin{equation}
n=\frac{\sqrt{(1-\epsilon\, y_+) (1+3 \epsilon \,y_+)}}{\sqrt{3}}
\label{eq:nspecial}
\end{equation}
with $\epsilon^2=1$. For $\epsilon=-1$, we need to restrict $y_+\leq1/3$, but this implies that $G(\eta)$ would have an additional root smaller than unity. We are thus left with $\epsilon=1$ to avoid any conical singularities, which also means $0<y_+\leq1$. The solution with $y_+=1$ has $n=0, \ G(\eta) = 1+\eta$, and corresponds to pure AdS$_4$.

Following \cite{deHaro:2000vlm} one can compute the resulting holographic stress energy tensor analytically, and fix the conformal frame by demanding that the boundary metric has fixed $g^{tt}_{\partial}=-1$. To accomplish this, we first change to Fefferman-Graham coordinates via the following asymptotic expansion
\begin{align}
&\eta =\frac{y_+ z}{r} \left[1-\frac{\left(2-5 n^2\right)}{2}\frac{z^2}{r^2}+\frac{\left(1-y_+\right) \left(1-12 y_+^2\right)}{9}\frac{z^3}{r^3}+\mathcal{O}(z^4)\right]\,,
\\
&\rho = \frac{1}{r}\left[1-\frac{1}{2}\frac{z^2}{r^2}+\frac{3 \left(1-n^2\right)}{8}\frac{z^4}{r^4}+\mathcal{O}(z^5)\right]\,,
\end{align}
which brings the metric into the following asymptotic form
\begin{equation}
\mathrm{d}s^2=\frac{L^2}{z^2}\left[\mathrm{d}z^2 + \mathrm{d}s^2_\partial +z^2 \mathrm{d}s^2_2+z^3\mathrm{d}s^2_3+\mathcal{O}(z^4)\right]\,.
\end{equation}
In the above expression we have
\begin{subequations}
\begin{equation}
\mathrm{d}s^2_\partial=-\dd t^2+\dd r^2+r^2 \left(\dd\phi - \frac{2\,n}{r} \dd t\right)^2
\label{eq:profilespecial}
\end{equation}
and
\begin{equation}
\mathrm{d}s^2_2=-\frac{\left(4-15 n^2\right) \mathrm{d}t^2}{10 r^2}-\frac{3 n^2}{2 r^2}\mathrm{d}r^2+\frac{5 n^2}{2} \left[\mathrm{d}\phi-\frac{2\left(1-5 n^2\right)}{5 n r}\mathrm{d}t\right]^2\,.
\end{equation}
\end{subequations}

Eq.~(\ref{eq:profilespecial}) allows us to identify $\omega(r)=2n/r$ for this particular profile, and thus $a=2\,n$, with $n$ given by Eq.~(\ref{eq:nspecial}).  
Since $0<y_+\leq1$,   $a$ has a maximum value at $y_+ = 1/3$ which corresponds to $a_{\max} = 4/3$. For $a=a_{\max}$, the solution exhibits no bulk curvature singularity, but we believe this is due to some special fine tuning induced by this very special profile. It is significant that $a_{\max} > 1$. For $a=1$, $\partial_t$ is null everywhere on the boundary and for $a>1$, it is spacelike. Thus the bulk solution develops an ergoregion 
before reaching $a_{\max}$. $a=1$ is reached when $y_+=1/3+\sqrt{7}/6$, corresponding to $n=1/2$.

\begin{figure}
\centering
    \includegraphics[width=0.45\textwidth]{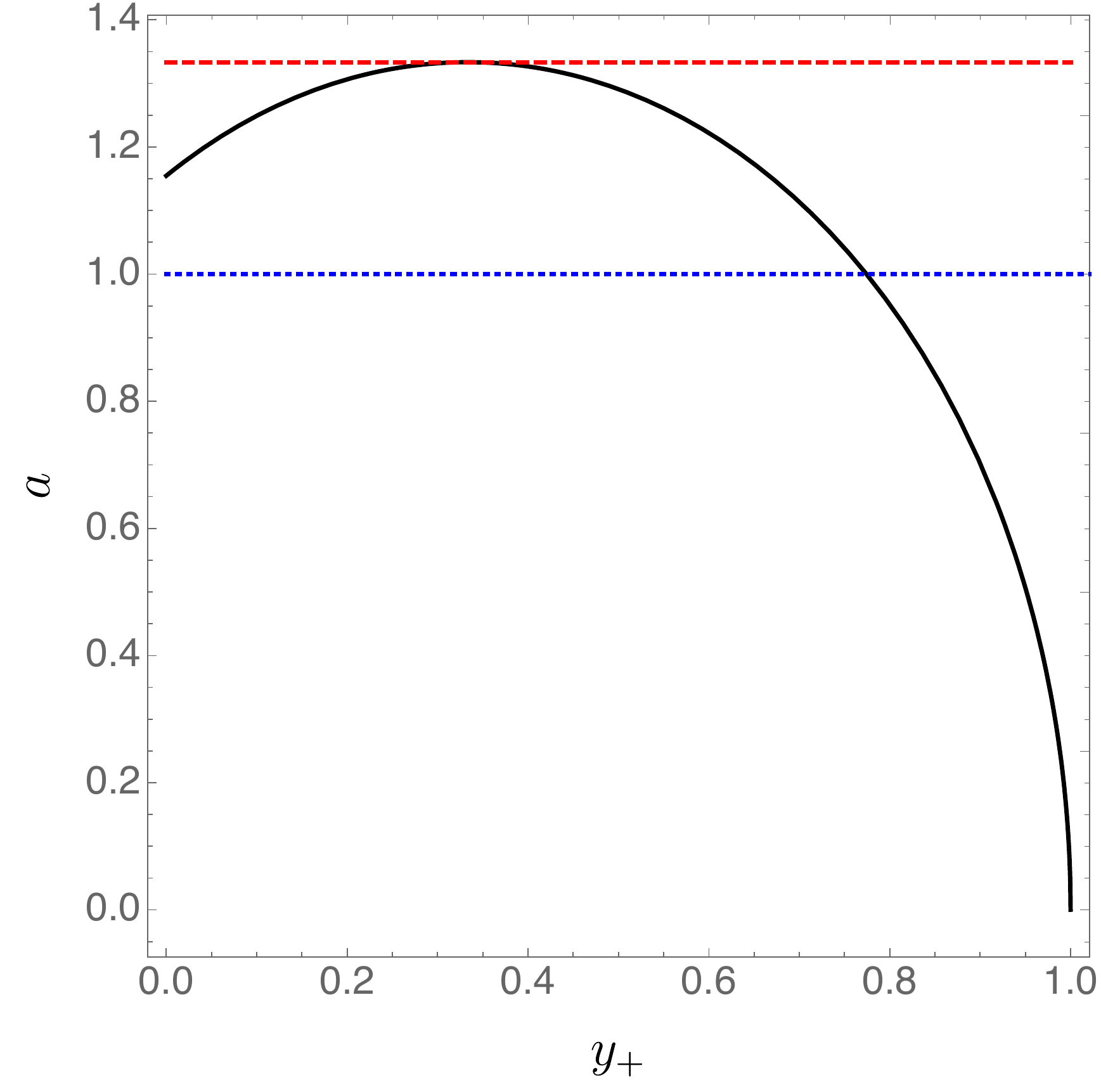}
  \caption{$a$ as a function of $y_+$: the horizontal red dashed like corresponds to $a=a_{\max}=4/3$ and the vertical blue dotted line to $a=1$, 
  beyond which $g_{tt}$ becomes everywhere spacelike at the boundary.}
\label{fig:special}
\end{figure}

As $y_+$ increases from zero to one, $a$ does not change monotonically. In Fig.~\ref{fig:special} we plot $a$ as a function of $y_+$ and mark both the onset of the ergoregion, $a=1$, and $a=a_{\max}$.

\begin{figure}
\centering
    \includegraphics[width=0.45\textwidth]{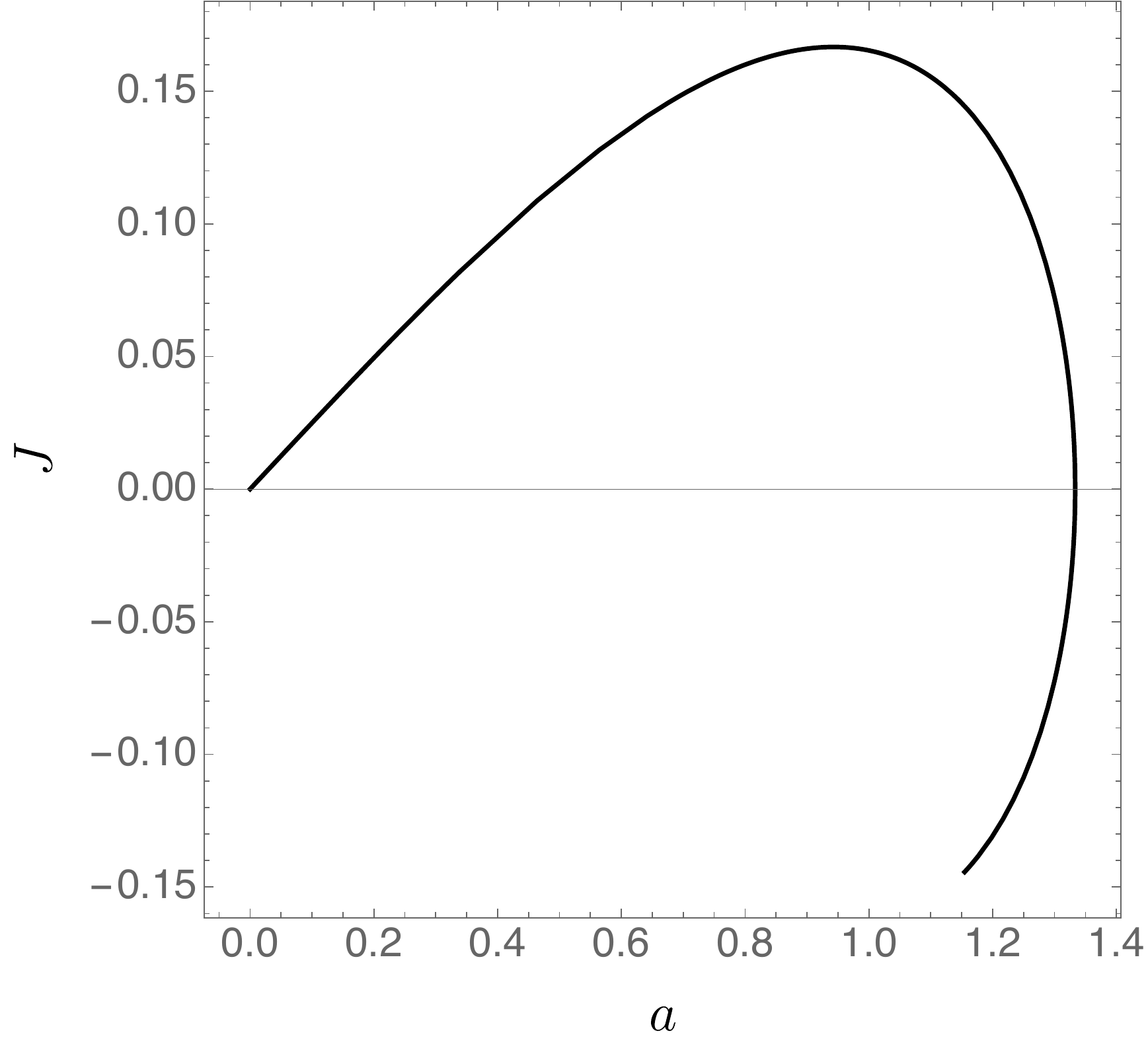}
  \caption{$J$ as a function of $a$, with $a=a_{\max}=4/3$ corresponding to a change of sign in $J$.}
\label{fig:angular}
\end{figure}

The holographic stress energy tensor is given in terms of $\mathrm{d}s^2_3$ by
\begin{equation}
\langle T_{\mu\nu}\mathrm{d}x^\mu\mathrm{d}x^\nu\rangle =\frac{3}{16 \pi}\mathrm{d}s^2_3= -\frac{\left(1-y_+\right) \left(1-12 y_+^2\right)}{24\pi r^3}\left[-\mathrm{d}t^2+\mathrm{d}r^2-2r^2\left(\mathrm{d}\phi-\frac{2\,n}{r}\mathrm{d}t\right)^2\right]\,,
\label{eq:stressspinning}
\end{equation}
where Greek indices run over the boundary spacetime directions.  It is easy to see that the angular momentum density, which is proportional to ${{T^t}_\phi}$, vanishes for all $r > 0$, so  it might at first seem that the total angular momentum will be zero. However, in deriving \ref{eq:stressspinning} we have completely neglected the fact that $g_{t\phi}$ is singular at $r=0$. If we were to take that into account, \ref{eq:stressspinning} would have a $\delta(r)$ contribution to  the angular momentum density. Instead of keeping track of this contribution, we  use the methods developed in \cite{Magnon:1985sc} to compute the total angular momentum. It turns out to be given by:
\begin{equation}
J = -\frac{\left(1-3 y_+\right) \sqrt{\left(2-3 y_+\right) y_++1}}{4 \sqrt{3}}\,.
\label{eq:momentum}
\end{equation}
Note that $J\geq0$  only for $y_+\in[1/3,1]$ and becomes negative for smaller $y_+$. This change in sign occurs precisely at $a_{\max}$. (See Fig.~\ref{fig:angular} for a plot of $J$ as a function of $a$.)  At present we have no understanding of why this is the case.
This dependence of $J$ on the amplitude turns out to be independent of the details of the differential rotation $\omega(r)$, and only depends on the fact that $\omega = a/r$ asymptotically.
We will  recover Fig.~\ref{fig:angular} (for $J>0$ ) when we construct zero-temperature solutions whose profile decays at large $r$ as $1/r$, but is regular as $r\to0$. In fact, the entire solution \eqref{eq:taubnut} is universal  in the sense that it provides the near horizon geometry for this class of profiles.
\subsubsection{\label{sec:irr}Metric \emph{ansatz} }
We first discuss the metric ansatz for profiles that fall off faster than $1/r$ asymptotically, which correspond to irrelevant deformations of the boundary. These solutions should have standard Poincar\'e horizons in the IR. We will use the coordinates first described in \cite{Horowitz:2014gva} and recently used in \cite{Crisford:2017zpi,Crisford:2017gsb}. We start with AdS written in Poincar\'e coordinates
\begin{equation}
\mathrm{d}s^2=\frac{L^2}{z^2}\left(-\mathrm{d}t^2+\mathrm{d}z^2+\mathrm{d}r^2+r^2\mathrm{d}\phi^2\right)\,,
\label{eq:poincare}
\end{equation}
where $z\in[0,+\infty)$, with $z=0$ marking the location of the conformal boundary, and $z=+\infty$ the Poincar\'e horizon. Furthermore, $(r,\phi)$ are standard polar coordinates in $\mathbb{R}^2$, with $r=0$ marking the axis of rotation. We now introduce two new coordinates $(\tilde{x},\tilde{y})$ which compactify both $r$ and $z$ in the following form
\begin{equation}
(z,r)=\frac{\tilde{y}\sqrt{2-\tilde{y}^2}}{1-\tilde{y}^2}(1-\tilde{x}^2,\tilde{x}\sqrt{2-\tilde{x}^2})\,,
\end{equation}
with $(\tilde{x},\tilde{y})\in[0,1]^2$. In terms of the $(t,\tilde{x},\tilde{y},\phi)$ coordinates, the metric on Poincar\'e AdS reads
\begin{equation}
\mathrm{d}s^2=\frac{L^2}{(1-\tilde{x}^2)}\left[-\frac{\left(1-\tilde{y}^2\right)^2}{\tilde{y}^2 \left(2-\tilde{y}^2\right)}\mathrm{d}t^2+\frac{4 \mathrm{d}\tilde{y}^2}{\tilde{y}^2 \left(1-\tilde{y}^2\right)^2 \left(2-\tilde{y}^2\right)^2}+\frac{4 \mathrm{d}\tilde{x}^2}{2-\tilde{x}^2}+\tilde{x}^2(2-\tilde{x}^2)\mathrm{d}\phi^2\right]\,.
\end{equation}
The Poincar\'e horizon is now at $\tilde{y}=1$, where the above metric reveals an AdS$_2$ like throat characteristic of zero temperature horizons. The boundary is located at $\tilde{x}=1$ and $\tilde{y}=0$
($\tilde{y}=0$ just corresponds to the point $z = r= 0$), and $\tilde{x}=0$ marks the axis of rotation. We note that, at the boundary, the relation between $\tilde{y}$ and $r$, reduces to
\begin{equation}
r=\frac{\tilde{y}\sqrt{2-\tilde{y}^2}}{1-\tilde{y}^2}\,.
\label{eq:rboundary}
\end{equation}

In order to use the DeTurck method we need to write down the most general line element compatible with our symmetries. Recall that our boundary metric has two commuting Killing fields, $\partial_t$ and $\partial_\phi$ and an additional discrete symmetry $(t,\phi)\to-(t,\phi)$. We assume that these symmetries extend smoothly into the bulk. The most general line element compatible with general diffeomorphisms along the $(\tilde{x},\tilde{y})$ directions takes the following form
\begin{multline}
\mathrm{d}s^2=\frac{L^2}{(1-\tilde{x}^2)^2}\Bigg[-\frac{\left(1-\tilde{y}^2\right)^2}{\tilde{y}^2 \left(2-\tilde{y}^2\right)}\,q_1\mathrm{d}t^2+\frac{4\,q_2\mathrm{d}\tilde{y}^2}{\tilde{y}^2 \left(1-\tilde{y}^2\right)^2 \left(2-\tilde{y}^2\right)^2}\\+\frac{4\,q_4}{2-\tilde{x}^2}\left(\mathrm{d}\tilde{x}+\frac{q_3}{1-\tilde{y}^2}\mathrm{d}\tilde{y}\right)^2+\tilde{x}^2(2-\tilde{x}^2)\,q_5(\mathrm{d}\phi-q_6 \mathrm{d}t)^2\Bigg]\,,
\label{eq:ansatz}
\end{multline}
where $q_i$, with $i\in\{1,\ldots,6\}$, are the functions of $(\tilde{x},\tilde{y})$ we wish to determine.

Four our reference metric we take
\begin{equation}
q_1=q_2=q_4=q_5=1\,,\quad q_3 = 0\,,\quad \text{and}\quad q_6=g(\tilde{y})\,,
\label{eq:ref0}
\end{equation}
where $g(\tilde{y})$ will control our chosen boundary profile for the differential rotation.

The boundary conditions are now easily obtained by requiring regularity at $\tilde{x}=0$, which in turn implies
\be
\partial_{\tilde{x}} q_1=\partial_{\tilde{x}} q_2= q_3=\partial_{\tilde{x}} q_4=\partial_{\tilde{x}} q_5=\partial_{\tilde{x}} q_6=0\,,\quad\text{and}\quad q_4=q_5\,.
\ee
At the conformal boundary, that is to say, at $\tilde{y}=0$ and $\tilde{x}=1$ we demand
\be
q_1=q_2=q_4=q_5=1\,,\quad q_3 = 0\,,\quad \text{and}\quad q_6=g(\tilde{y})\,,
\ee
and finally, since we expect these solutions to have a standard Poincar\'e horizon, at $\tilde{y}=1$ we have
\be
q_1=q_2=q_4=q_5=1\,,\quad q_3 = 0\,,\quad \text{and}\quad q_6=0\,.
\ee
Note that consistency of our boundary conditions imposes $g(1)=0$.

The case of marginal deformations ($\omega \sim 1/r$) is different, since we expect the IR to be deformed away from pure AdS into the family of exact solutions discussed in section \ref{sec:spinning}. The metric \emph{ansatz} remains as in Eq.~\ref{eq:ansatz} except we set $q_6 = (1-\tilde{y}^2) \hat{q}_6$, and express all boundary conditions in terms of $\hat{q}_6$. Note that this means at the boundary (both at $\tilde{x}=1$ and $\tilde{y}=0$) we want $\hat{q}_6=a$, and that at the symmetry axis we still have $\partial_{\tilde{x}} \hat{q}_6=0$. The only significant change comes at the would be Poincar\'e horizon, where we impose
\begin{equation}
q_1=q_2\,,\quad \partial_{\tilde{y}} q_1=\partial_{\tilde{y}} q_2=\partial_{\tilde{y}} q_4=\partial_{\tilde{y}} q_5=\partial_{\tilde{y}} \hat{q}_6=q_3=0\,,
\label{eq:bcsmarginal}
\end{equation}
which are enforced via regularity in ingoing Eddington-Finkelstein coordinates.

\subsubsection{Results}
\label{sec:resultsT0}

We will first focus on the following class of profiles for the differential rotation:
\begin{equation}
\omega(r)=\frac{A}{\displaystyle\left(1+\frac{r^2}{\sigma^2}\right)^{n/2}}\,,
\label{eq:pro1}
\end{equation}
where $A$ is the boundary profile amplitude, $\sigma$ is a length scale, and $n$ is a positive integer\footnote{This should not be confused with the $n$ which appeared in section \ref{sec:spinning} which will not be referred to again.}. At large $r$, these profiles decay like $1/r^n$.  Since the boundary metric \eqref{eq:bndmetric} is only determined up to conformal rescalings,  we can always rescale $A$ and $\sigma$ such that the only meaningful quantity is $a\equiv A\sigma$. We will fix $\sigma=1$ in the numerics, and so $A=a$.  In terms of the $\tilde{y}$ coordinate (\ref{eq:rboundary}), we have
\begin{equation}
\omega(r)=  \frac{a}{(1+r^2)^{n/2}} = g(\tilde{y})=a\,(1-\tilde{y}^2)^n\,.
\label{eq:pro1a}
\end{equation}

\begin{figure}
\centering
    \includegraphics[width=0.45\textwidth]{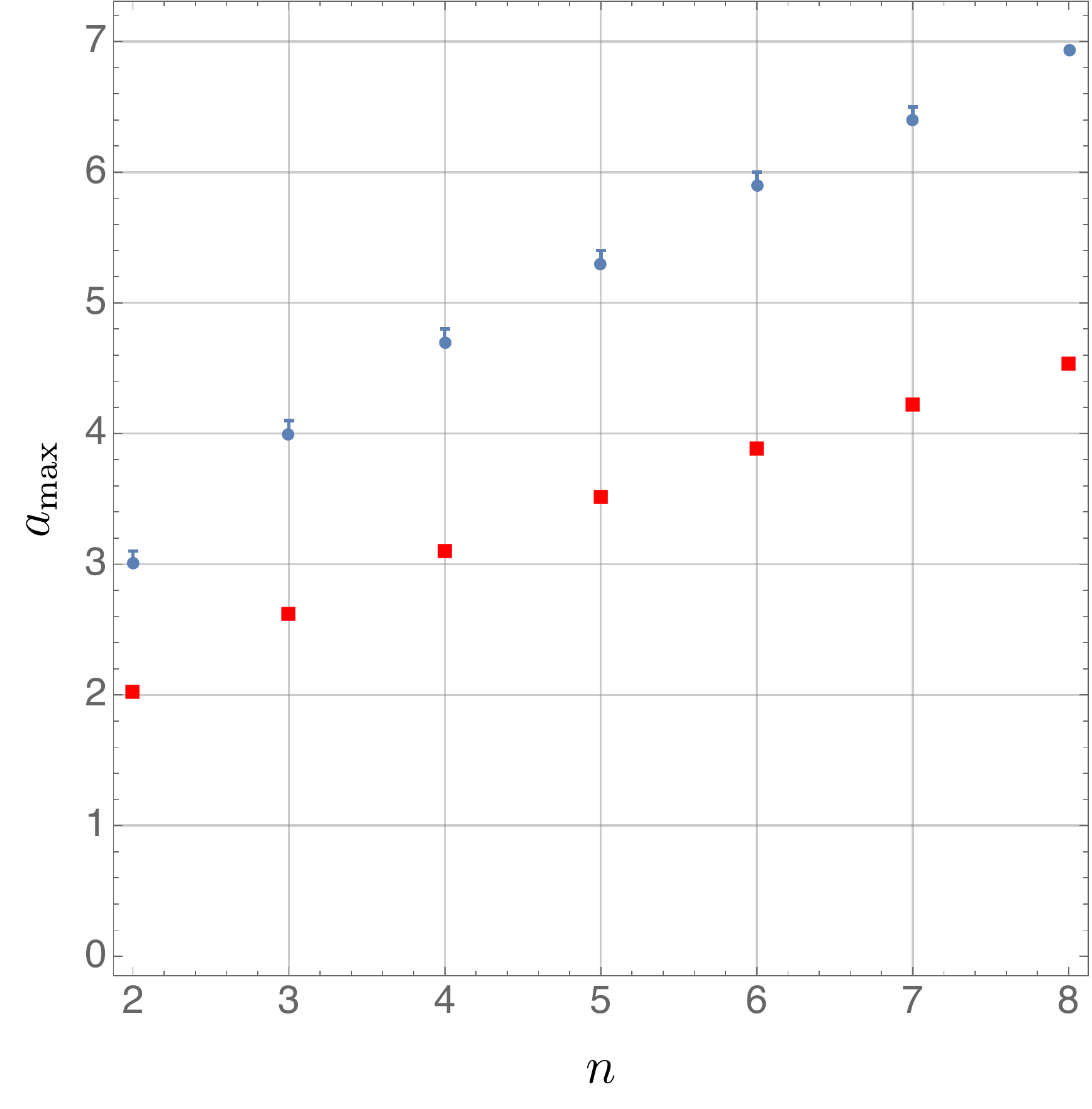}
  \caption{$a_{\max}$ (represented by the blue disks) and $a_{ergo}$ (represented by the red squares) as a function of $n$. The error bars in determining $a_{\max}$ are computed via the failure of our code to find solutions for the upper range of $a$.}
\label{fig:amaxpro1}
\end{figure}

We start with results for $n>1$.  (The special case $n=1$ will be discussed shortly.) For each fixed value of $n$, we construct the solutions numerically by increasing $a$ starting with $a=0$. In all cases we find  a critical value $a=a_{\max}$, at which the solution becomes singular.  This maximum amplitude  always  lies past the point where an ergoregion develops on the boundary, $a_{\ergo}$. For $2 \leq n \leq 8$, this is shown in Fig.~\ref{fig:amaxpro1} where we have plotted both $a=a_{\max}$ (represented by the blue disks with error bars\footnote{The error bars in determining $a_{\max}$ simply reflect the fact that a solution exists at the blue dot but not at the upper end of the error bar (using a uniform grid over $a$). In the following, when we quote results for $a_{\max}$, we mean this largest value for which we have found a solution, and not literally the singular solution.}) and $a_{\ergo}$ (represented by the red squares) for various profiles.   
 All of these solutions have a standard Poincar\'e horizon as expected. 
For the above class of profiles, $a_{\ergo}$ is given by
\be
a_{\ergo}=\sqrt{n} \left(\frac{n}{n-1}\right)^{\frac{n-1}{2}}\,.
\ee
For $a> a_{\ergo}$, the ergoregion is an annular region around the origin on the boundary, and extends into the bulk. For $a= a_{\ergo}$ the ergoregion collapses to a single circle and has been called an evanescent ergoregion \cite{Gibbons:2013tqa}. It does not extend into the bulk.  Note that both $a_{\max}$ and $a_{\ergo}$ increase with $n$. So when the differential rotation on the boundary falls off faster, solutions exist for a larger amplitude. It is also clear from Fig.~\ref{fig:amaxpro1} that the difference $a_{\max}-a_{\ergo}$ increases with $n$.

\begin{figure}
\centering
    \includegraphics[width=0.45\textwidth]{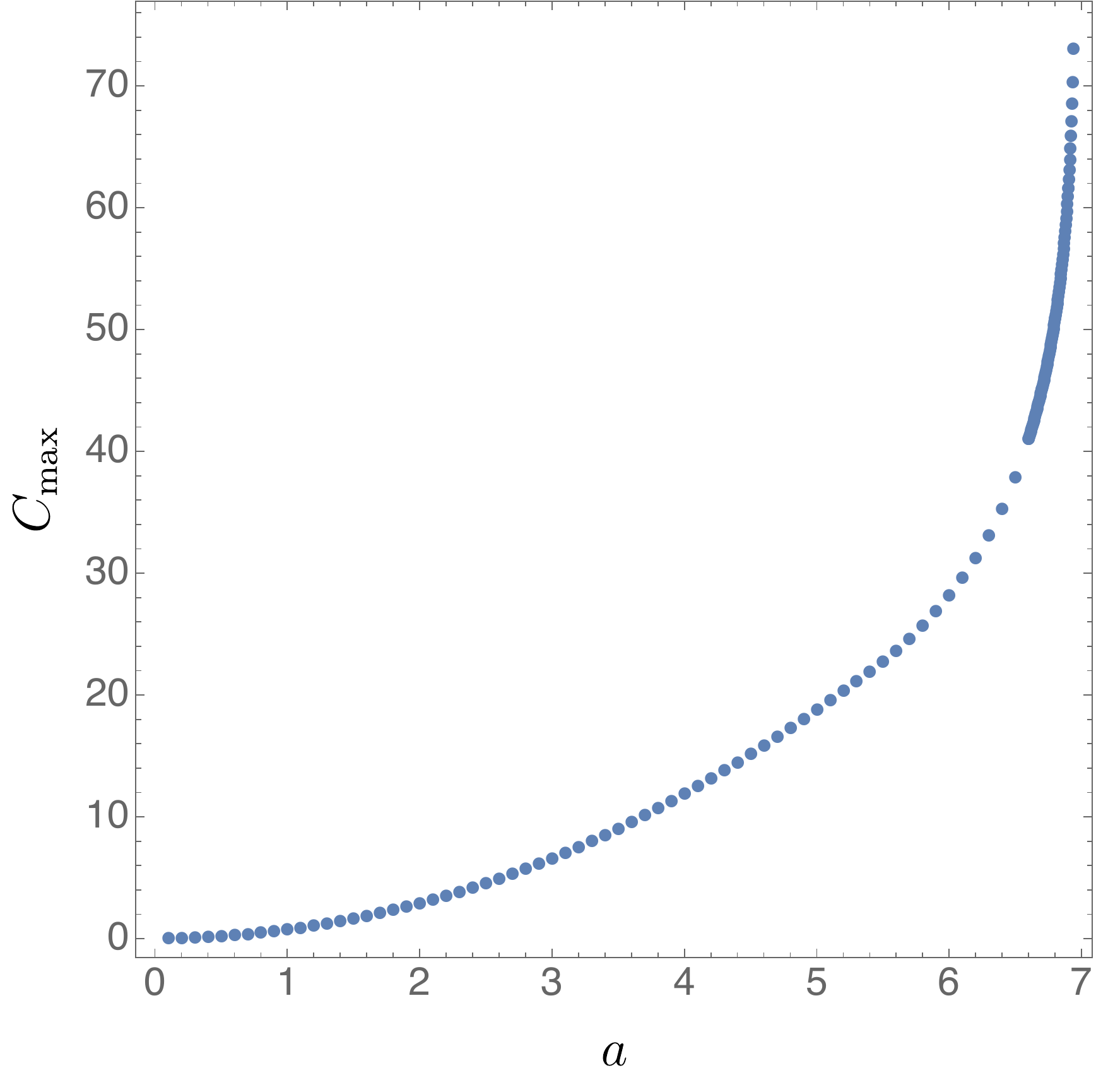}
  \caption{Maximum value of $C_{abcd}C^{abcd}$ over spacetime, computed with $n=8$, as a function of $a$.
  }
\label{fig:C2max}
\end{figure}

\begin{figure}
\centering
    \includegraphics[width=0.9\textwidth]{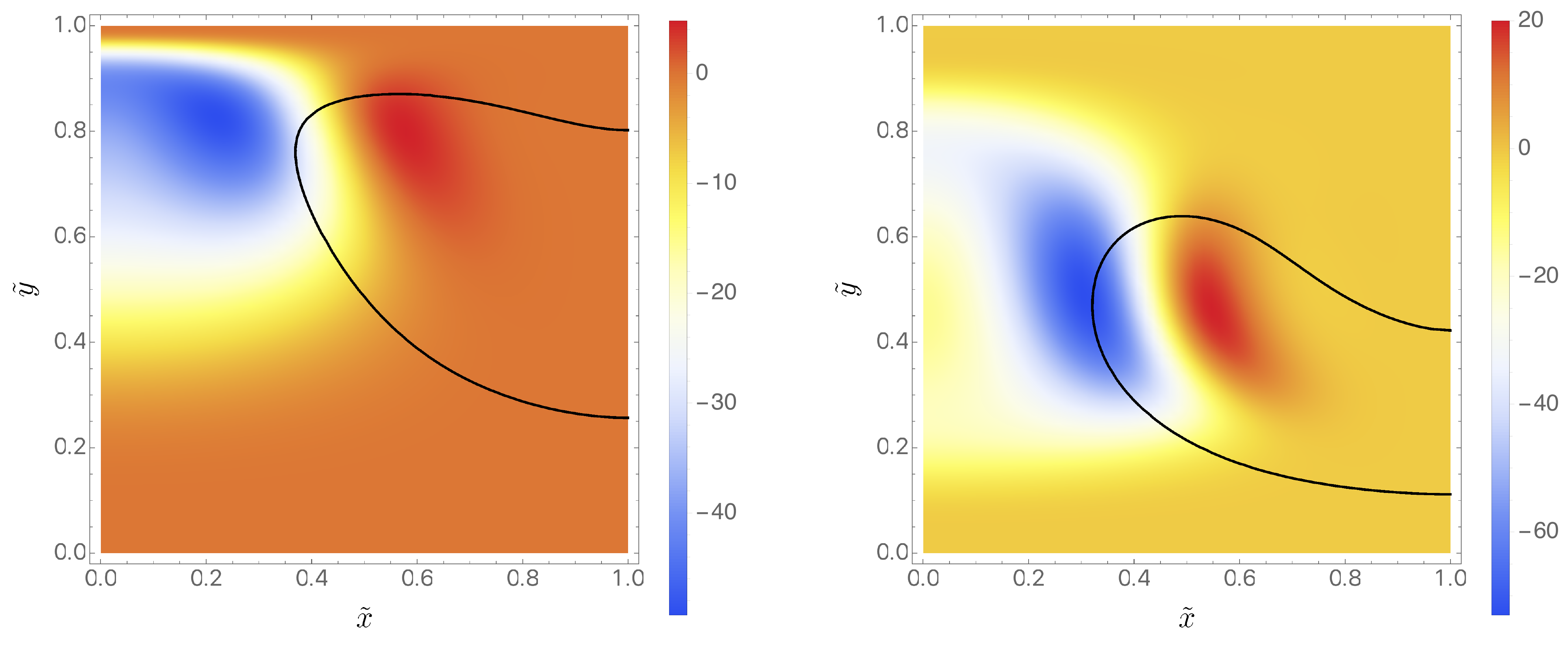}
  \caption{The Weyl tensor squared computed with $a=a_{\max}$ for $n=2$ (left panel) and $n=8$ (right panel). Note that the maximum curvature does not appear on the axis ($\tilde{x}=0$). The ergoregion lies inside the black line, and does not always contain the maximum curvature. (In these coordinates, the asymptotic boundary is $\tilde{x} =1$.)}
\label{fig:weyl}
\end{figure}

To show the formation of a singularity as we approach $a_{\max}$, we monitor the square of the Weyl tensor $C_{abcd}C^{abcd}$ throughout spacetime. Let
\be\label{eq:cmax}
C_{\max}\equiv \underset{\mathcal{M}}{\max}\left|C_{abcd}C^{abcd}\right|\,,
\ee
where $\mathcal{M}$ denotes our spacetime manifold. In Fig.~\ref{fig:C2max} we plot  $C_{\max}$ as a function of $a$ for $n=8$. The rapid growth as $a\to a_{\max}$ is clearly visible. To gain more information about where the singularity appears,
in Fig.~\ref{fig:weyl} we plot $C_{abcd}C^{abcd}$ for $a=a_{\max}$ and $n=2$ (left panel) and $n=8$ (right panel). Since the rotation axis corresponds to $\tilde{x}=0$, it is clear that the large curvature is occurring away from this axis. One might wonder if it always occurs inside the ergoregion. To check this, we have denoted the boundary of the ergoregion by a solid black line  in Fig.~\ref{fig:weyl}.   It is clear that the maximum curvature is not always inside the ergoregion.

We next investigate physical quantities like the energy density $\rho$ and angular momentum density $j$. These are defined in terms of the holographic stress tensor by
\begin{subequations}
\begin{equation}
\rho \equiv -\langle T^{t}_{\phantom{t}t}\rangle\,,
\end{equation}
and 
\begin{equation}
j \equiv \langle T^{t}_{\phantom{t}\phi}\rangle\,.
\end{equation}
\label{eq:thermo}
\end{subequations}
Within our symmetry class, the holographic stress energy tensor has four non-zero components. In addition, it should be traceless and conserved, which gives two constraints amongst these four components. Thus, the full stress energy tensor is determined by  $\rho$ and $j$.
 
\begin{figure}
\centering
    \includegraphics[width=0.9\textwidth]{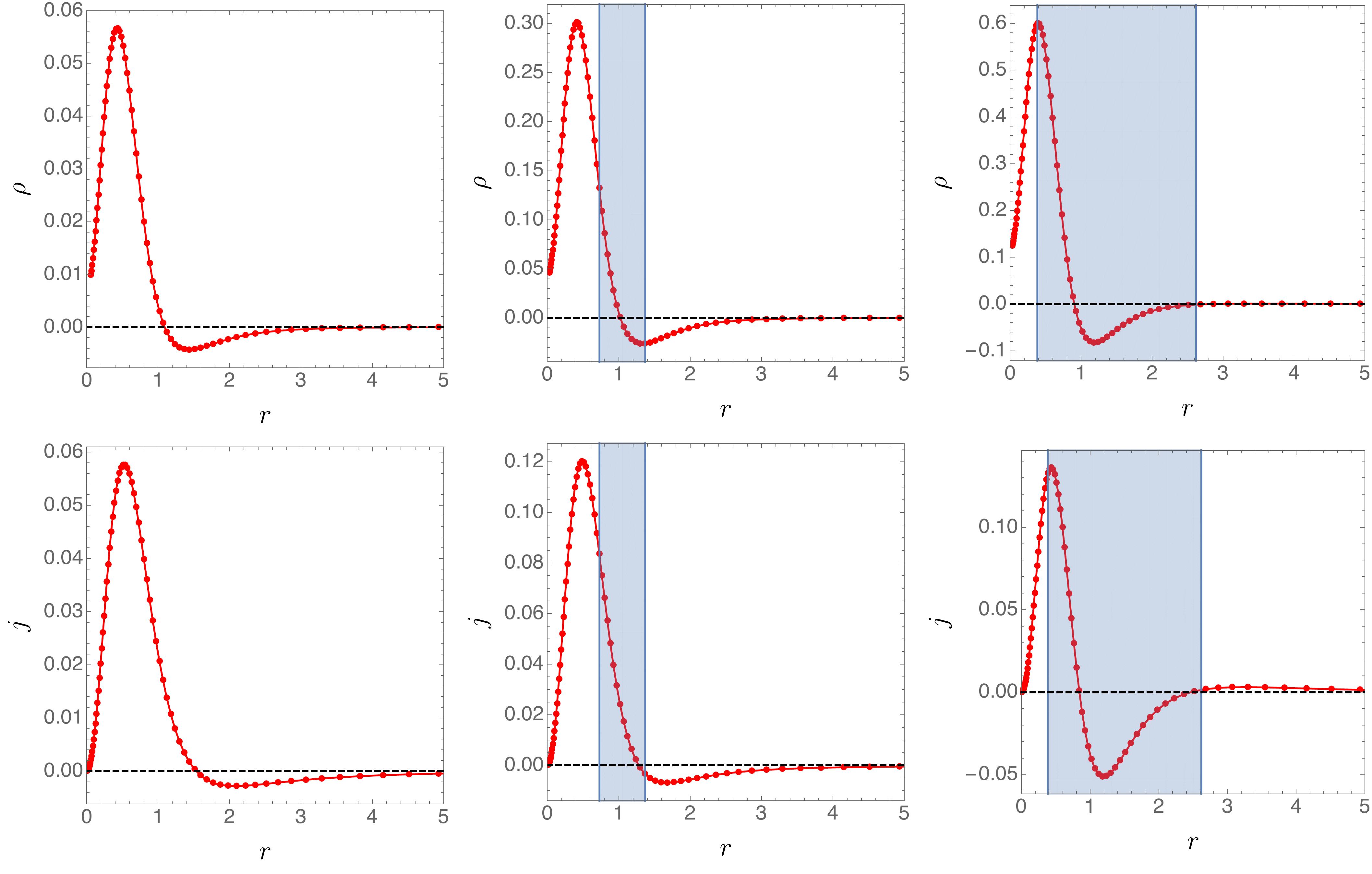}
  \caption{The holographic energy density (top row) and holographic angular momentum density (bottom row) computed with $n=2$. The dashed horizontal line marks $0$, and the 
  blue region indicates the location of the boundary ergoregion. From left to right, in each of the rows, we have $a=0.9,2.1,3$. (The last value corresponds to $a_{\max}$.)}
\label{fig:holo}
\end{figure}

 Two important questions are whether $\rho$ and $j$ change their behavior qualitatively after the ergoregion forms on the boundary, and whether  they diverge as we approach $a_{\max}$. To check this, we computed the holographic stress energy tensor using \cite{deHaro:2000vlm}, and following \emph{mutatis mutandis} section \ref{eq:holeplanar}. We find that the answer to both questions is no: the formation of the ergoregion does not dramatically affect these quantities and they appear to remain finite. This is illustrated in Fig.~\ref{fig:holo} for $n=2$ and several values of $a$, including $a = a_{\max}$ (most right column). Note that the scales on the vertical axis are different in the six plots, and the maximum values of $|\rho|$ and $|j|$ tend to increase with $a$. Curiously, although $\rho $ remains finite, for $a=a_{\max}$  it reaches a maximum precisely at the edge of the ergoregion. We see this happening for all values of $n$, and for different profiles. 
 Note that  even though we have imposed a differential rotation $\omega(r)$ that is positive everywhere, the induced angular momentum density $j$ takes both positive and negative values. In fact, the total angular momentum, $J$, in the spacetime turns out to be exactly zero (to machine precision). This is directly analogous to what was found in  \cite{Horowitz:2014gva} where static solutions of Einstein-Maxwell were discussed. There it was shown that a localized positive chemical potential that falls off faster than $1/r$ produces regions of both positive and negative  charge density, but the total charge remains exactly zero. 

\begin{figure}
\centering
    \includegraphics[width=0.45\textwidth]{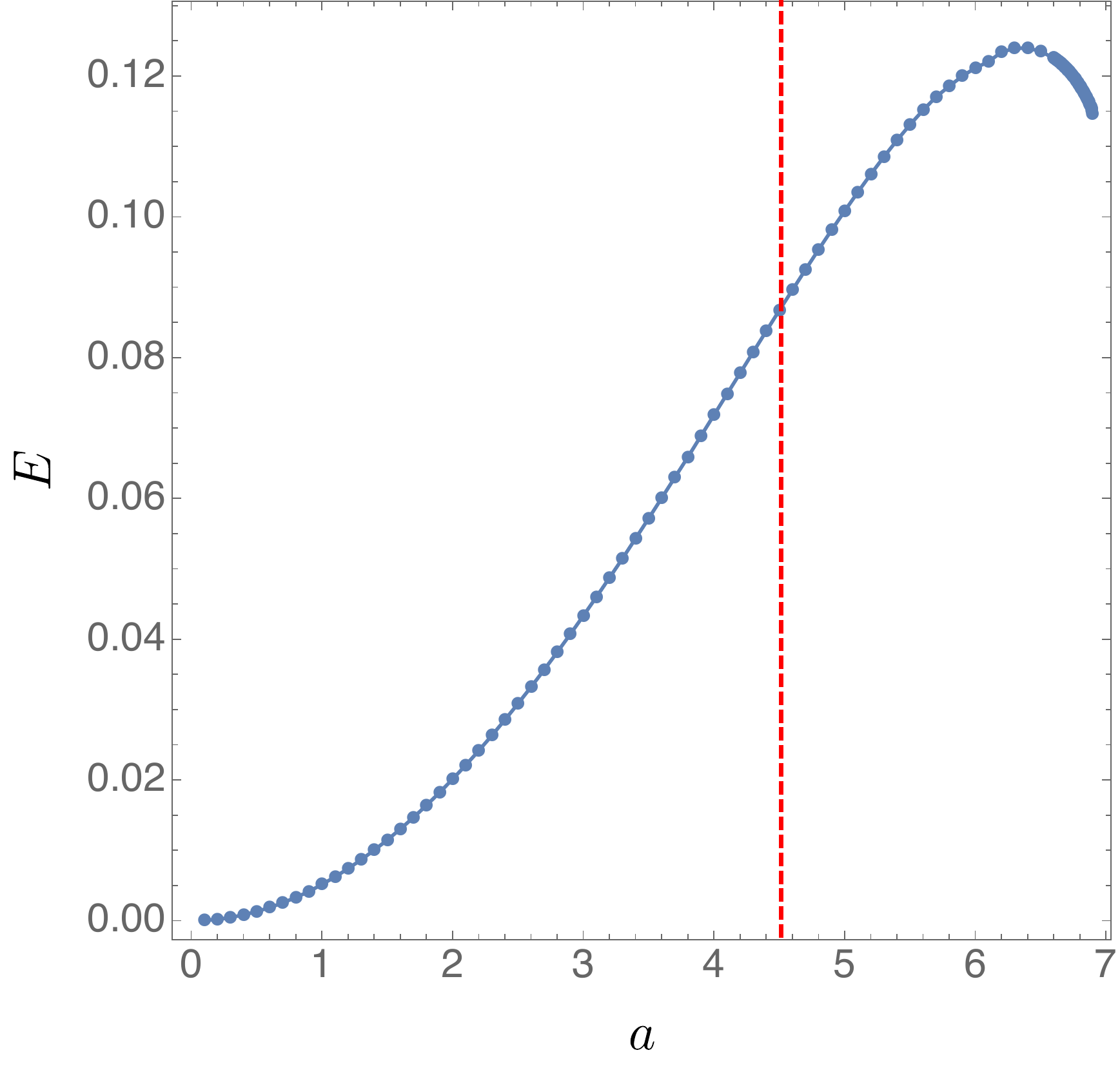}
  \caption{Total energy $E$ as a function of $a$, for $n=8$. The verticle dashed line corresponds to $a=a_{\ergo}$ and the curve ends at $a_{\max}$.}
\label{fig:energy}
\end{figure}

 The total energy is plotted in Fig.~\ref{fig:energy} for $n=8$. We note that the energy is always positive even after the formation of the boundary ergoregion, denoted by the vertical dashed line \emph{i.e.} $a=a_{\ergo}$.  The behaviour of the energy as we approach $a_{\max}$ is puzzling to us, \emph{i.e.} we do not understand why it is not monotonic with increasing $a$. 

It is clear from Fig.~(\ref{fig:amaxpro1})  that the difference $a_{\max} - a_{\ergo}$ depends on the profile for the differential rotation. It turns out that one can make this difference arbitrarily small by a judicious choice of profile\footnote{We will see in section \ref{sec:blacknon} that another way to make this difference arbitrarily small is to go to high temperature.}. One way to do this is to choose a profile that is very sharply peaked at the origin. Consider
\begin{equation}
\omega(r)=\frac{a}{\displaystyle\left(b^2+{r^2}\right)^{1/2}\left(1+{r^2}\right)^{1/2}}
\label{eq:pro2}
\end{equation}
We have added a new parameter $b$, which  controls the height and thickness of the profile around $r=0$. In terms of the coordinate (\ref{eq:rboundary}), this profile reads
\begin{equation}
\omega(r)=a\,\frac{(1-\tilde{y}^2)^2}{\left[b^2(1-\tilde{y}^2)^2+\tilde{y}^2(2-\tilde{y}^2)\right]^{1/2}}
\label{eq:pro2a}
\end{equation}

\begin{figure}
\centering
    \includegraphics[width=0.45\textwidth]{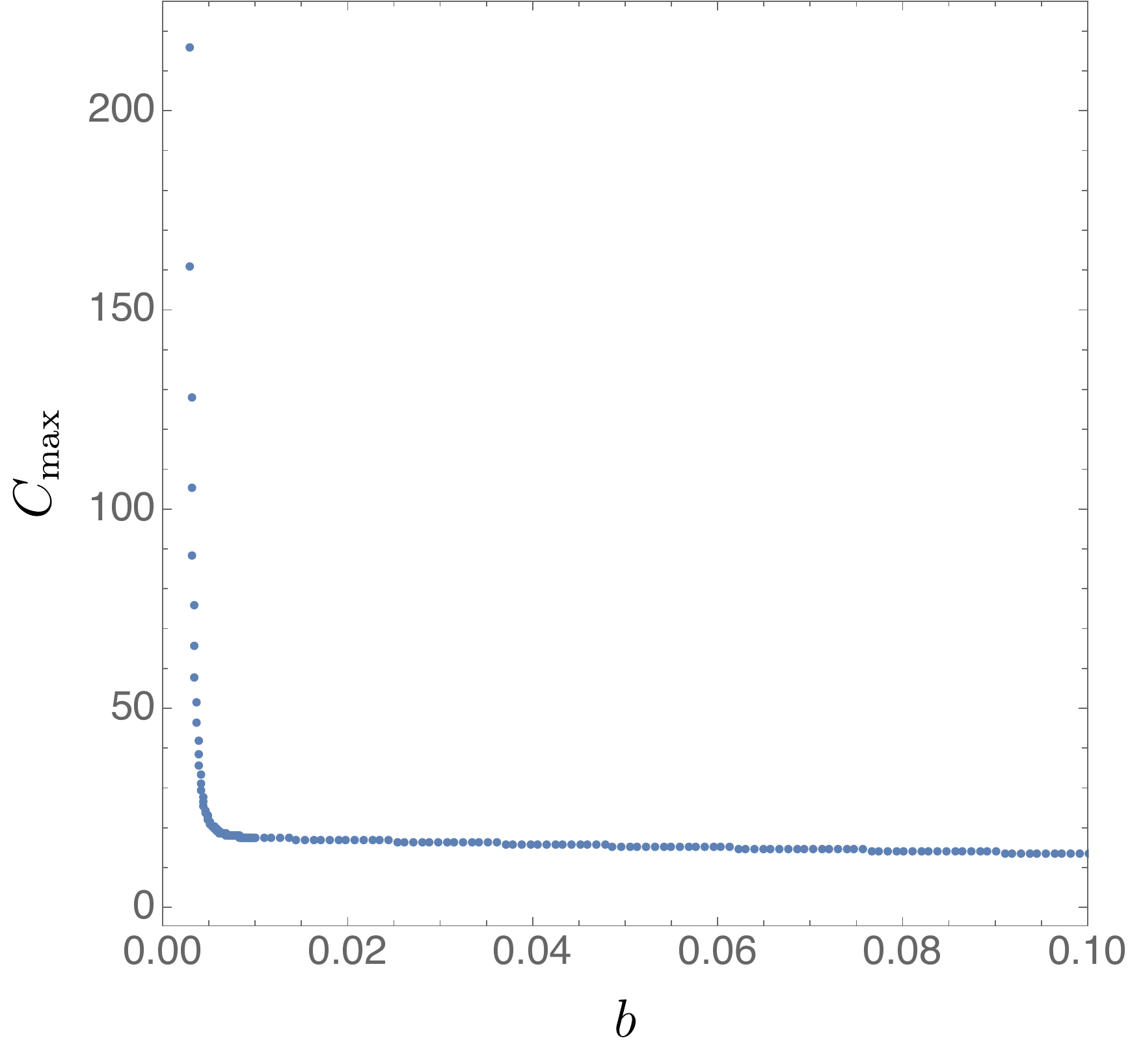}
  \caption{$C_{\max}$ as a function of $b$ for $a=1$, $n=2$ and profile \ref{eq:pro2}: the blow up as $b\to0$ suggests that the solution stops existing precisely when an ergoregion first forms.}
\label{fig:kretextreme}
\end{figure}

To see the effect of $b$,  we set $a=1$ and decrease $b$.  Then $g_{tt} = -1 + r^2 \omega^2(r) < 0 $ everywhere for $b>0$, but vanishes at $r=0$ when $b=0$. So the ergoregion first forms at the origin in this case. Note that for $b=0$, the profile looks like $a/r$ near the origin, just like the spinning top solution discussed in section \ref{sec:spinning}. We now compute the maximum curvature $C_{\max}$ \eqref{eq:cmax} as a function of $b$ to see when a singularity forms. The results are shown in  Fig.~\ref{fig:kretextreme},  where we see that solutions exist for all $b >0$, but the $b\to0$ limit appears to be singular. So for this profile, stationary solutions cease to exist precisely when the ergoregion first forms.
 
\begin{figure}
\centering
    \includegraphics[width=0.45\textwidth]{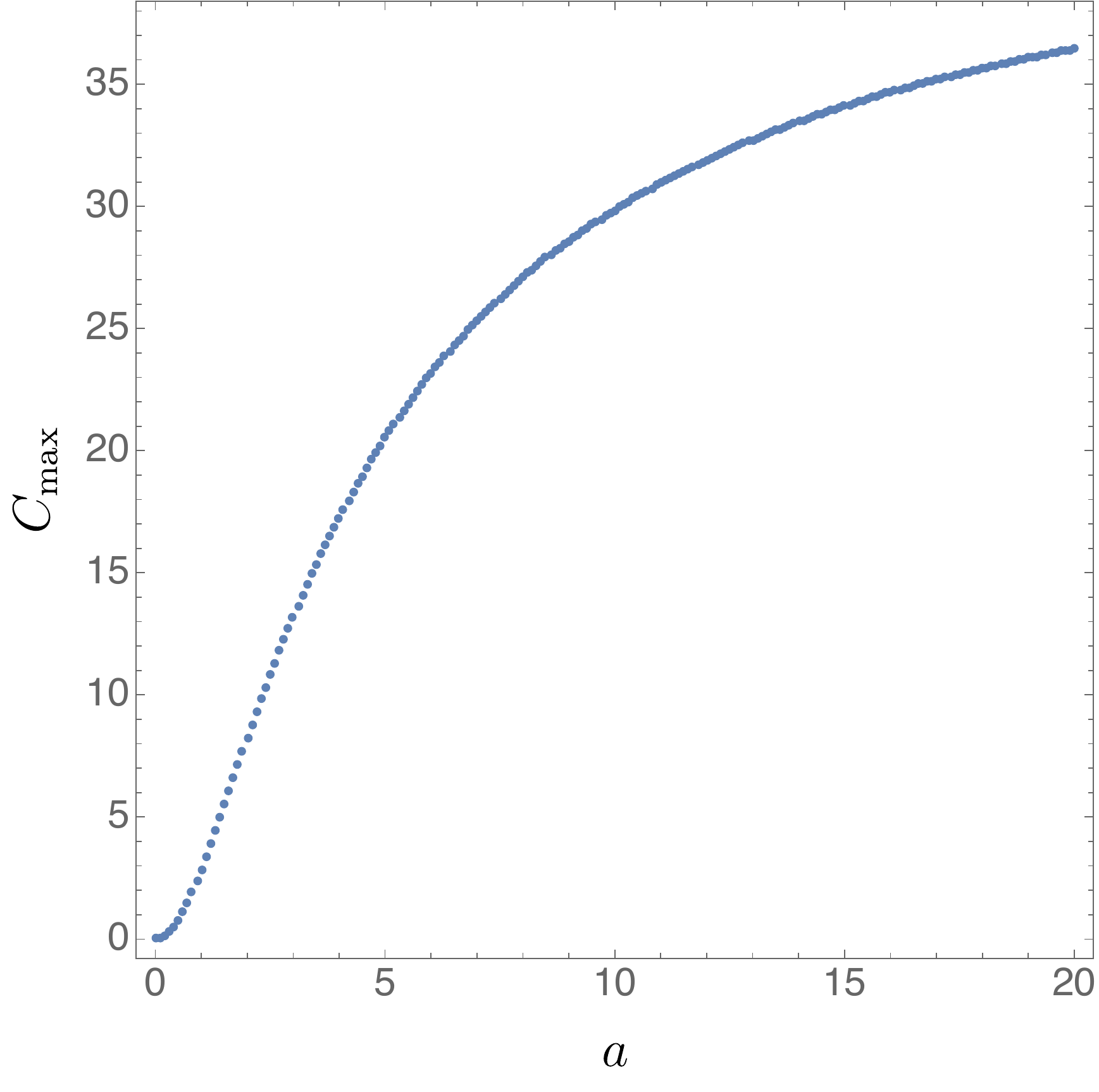}
  \caption{$C_{\max}$ as a function of $a$ for boundary metrics without ergoregions: there is no indication that the metric is becoming singular at any finite value of $a$.}
\label{fig:kret}
\end{figure}

To see if we could find situations where $a_{\max}$ is reached before an ergoregion exists, we considered a class of rotating boundary geometries without ergoregions:
\begin{equation}
\dd s^2_\partial = -\dd t^2+\dd r^2+r^2 \dd\phi^2 -2 r^2\omega(r) \dd t \dd\phi\;,
\label{eq:bndmetric2}
\end{equation}
These solutions can be constructed exactly as before (section \ref{sec:irr}), except that we change our reference metric to be such that no ergoregion is present. For the reference metric we take all functions as in Eq.~(\ref{eq:ref0}), except for $q_1$ which is now given by
\begin{equation}
q_1 = 1+g(\tilde{y})^2\frac{\tilde{y}^2(2-\tilde{y}^2)}{(1-\tilde{y}^2)^2}\,,
\end{equation}
The form of $\omega(r)$ is again given by \eqref{eq:pro1a}. We studied in great detail the cases with $n=2,4,6,8$ and we found no upper bound on $a$. In all these cases, we were able to reach $a\sim 20$ without seeing any indication that the solution is becoming singular. In Fig.~\ref{fig:kret} we plot $C_{\max}$ as a function of $a$ for $n=2$. Contrary to the case where the ergoregion is present, the maximum now occurs along the axis of rotation.

Finally, we  briefly discuss the case $n=1$.  There are two qualitative differences from the $n>1$ solutions. First, the IR geometry is not given by a standard Poincar\'e horizon, but rather by the extremal horizon of a member of the holographic  spinning top solution discussed in section \ref{sec:spinning}. In other words,  any profile that asymptotically behaves like $\omega(r) \sim a/r$ has the same IR geometry as the solution where $\omega(r) = a/r$ everywhere. Note that our boundary conditions (\ref{eq:bcsmarginal}) do not impose this as a Dirichlet condition. Instead, this emerges as the natural IR solution.

\begin{figure}
\centering
    \includegraphics[width=0.45\textwidth]{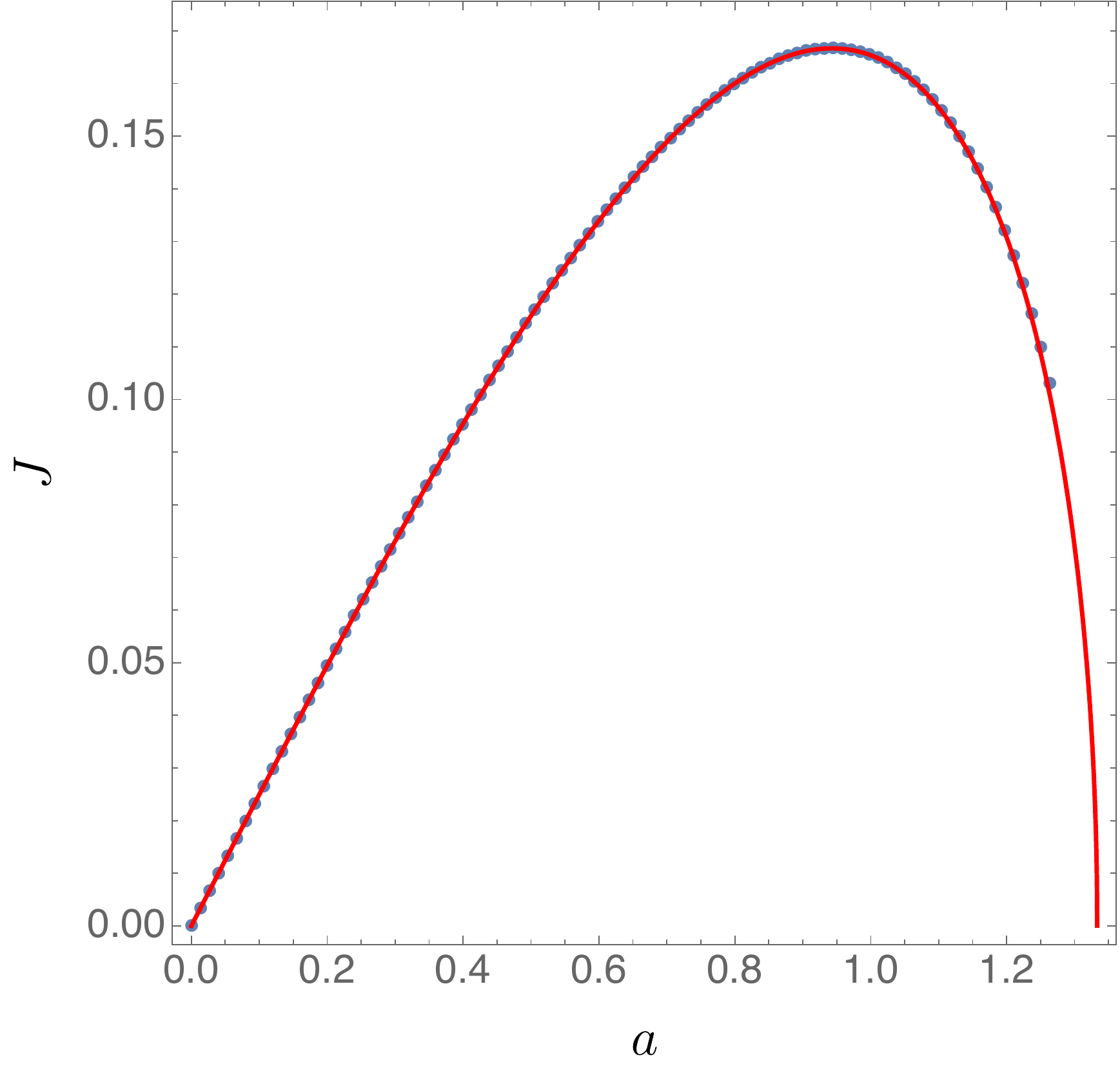}
  \caption{Total angular momentum as as a function of $a$, for the marginal case ($n=1$). The solid red line is the result for the spinning top shown in Fig.~\ref{fig:angular}.}
\label{fig:match}
\end{figure}

Second, the total angular momentum is no longer zero. Since our $n=1$ boundary profile
\be
\omega(r) = \frac{a}{\sqrt{1+r^2}}=a(1-\tilde{y}^2)\,,
\ee
is not singular at $r=0$, we can compute the total angular momentum just by integrating the angular momentum density. The result is depicted in Fig.~\ref{fig:match}, where we also superimpose our exact result (\ref{eq:momentum}).
We see that the angular momentum of the regular profile agrees with that of the spinning top with the same coefficient of the $1/r$ fall-off.

\subsubsection{Stability analysis}

For $a < a_{\ergo}$, we expect our solutions to be stable, but for $a > a_{\ergo}$, 
they may be unstable to nonaxisymmetric perturbations due to the   
superradiant instability. 
To investigate this, instead of studying the full gravitational perturbations, we will  study 
 perturbations by a  massless scalar field.   So we add a field $\Phi$ satisfying the massless wave equation 
\begin{equation}
\Box \Phi =0\,.
\label{eq:scalar}
\end{equation}
Since the background is stationary and axisymmetric, we can Fourier decompose $\Phi$ as
\begin{equation}
\Phi = e^{-i\,\omega\,t+i\,m\,\phi}\,\widehat{\Phi}(\tilde{x},\tilde{y})\,,
\label{eq:sepa}
\end{equation}
and find the quasinormal mode spectrum, \emph{i.e.} the complex values of $\omega$ for which $\widehat{\Phi}$ is normalisable at the conformal boundary, and regular at the horizon. We are primarily interested in finding the onset of the instability, which occurs for $\omega=0$. We can then interpret Eq.~(\ref{eq:scalar}) as an eigenvalue equation for $m^2$ for a given value of $a$. Of course, we want $m\in\mathbb{Z}$, since $\phi$ is chosen to have period $2\pi$.

We now have to discuss the thorny issue of boundary conditions. The best way to do this is to expand Eq.~(\ref{eq:scalar}) around each of our integration boundaries and use Frobenius's method to extract the leading non-analytic behaviour. For $\omega=0$, we find the following behaviour:
\be
\widehat{\Phi}\simeq \left(1-\tilde{x}^2\right)^3 \tilde{x}^{|m|} \left(2-\tilde{x}^2\right)^{|m|/2} \tilde{y}^{|m|+2 p+3} \left(2-\tilde{y}^2\right)^{\frac{1}{2} (|m|+2 p+3)}C_+\,,
\ee
where $p$ is an integer and $C_+$ is a smooth function around $\tilde{y}=0$. In terms of the original coordinates $(r,z)$ of the line element (\ref{eq:poincare}) this reads
\be
\widehat{\Phi}\simeq \frac{r^{|m|} z^3 \left(r^2+z^2\right)^p}{\left(1+r^2+z^2\right)^{\frac{1}{2} (3+|m|+2 p)}}\widehat{C}_+(r,z)\,,
\ee
where $\widehat{C}_+$ is a smooth function of $r^2+z^2$. Regularity at the origin thus demands $p=0$. Note that the factor $r^{|m|}$ is needed to cancel the non-analytic behaviour of $e^{im\phi}$ included in Eq.~(\ref{eq:sepa}). We thus perform the following change of variables
\be
\widehat{\Phi}= \left(1-\tilde{x}^2\right)^3 \tilde{x}^{|m|} \left(2-\tilde{x}^2\right)^{|m|/2} \tilde{y}^{|m|+3} \left(2-\tilde{y}^2\right)^{\frac{1}{2} (|m|+3)}\widetilde{\Phi}
\ee
and solve for $\widetilde{\Phi}$ numerically. All we are missing is a choice of boundary conditions. At $\tilde{x}=0$ and $\tilde{x}=1$ we find $\partial_{\tilde{x}} \widetilde{\Phi}=0$, while at $\tilde{y}=0$ we have $\partial_{\tilde{y}} \widetilde{\Phi}=0$. Finally at the Poincar\'e horizon we find $\widetilde{\Phi}=0$.

\begin{figure}
\centering
    \includegraphics[width=0.45\textwidth]{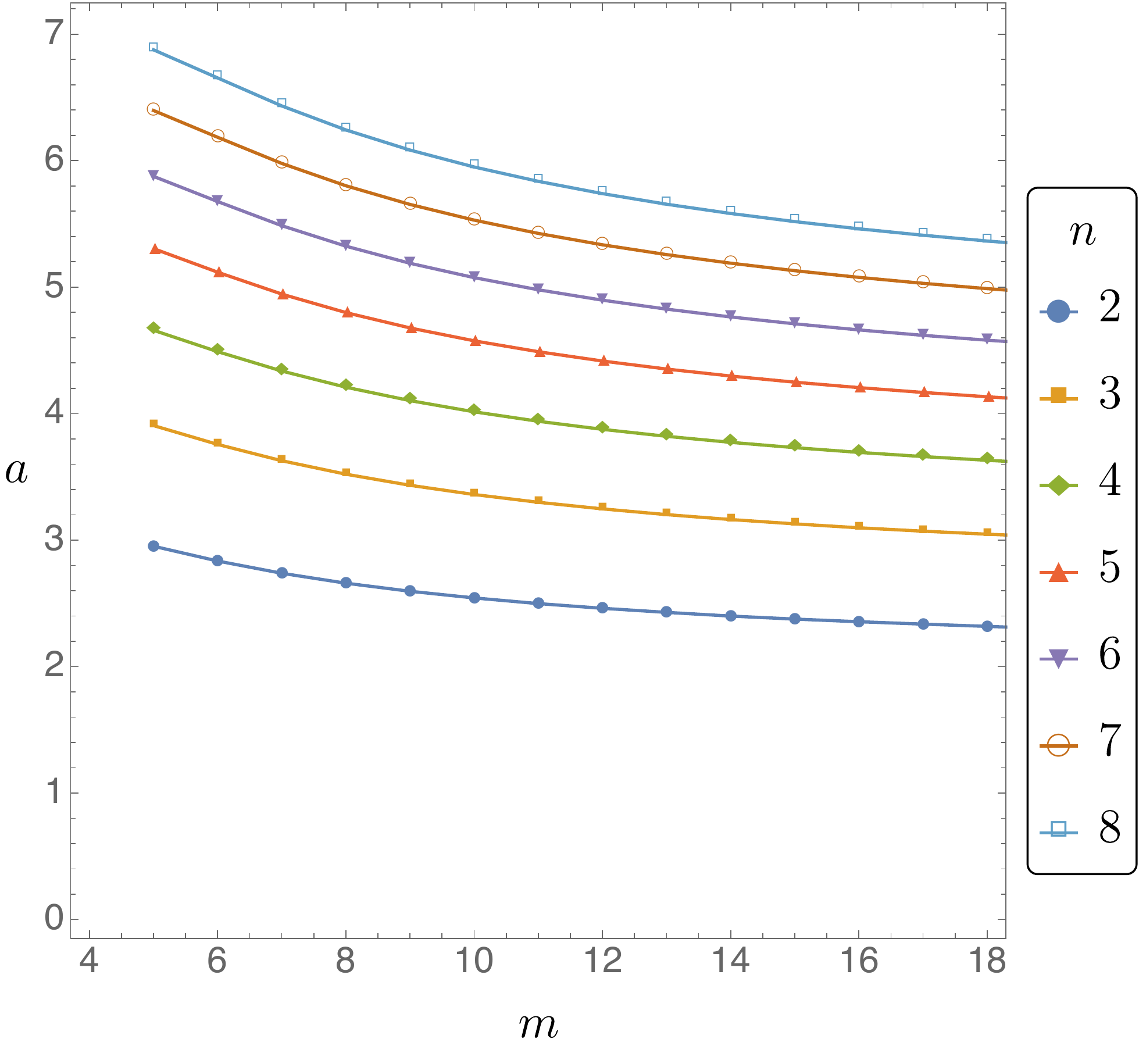}
  \caption{Onset of superradiance for each mode $m$, around our profile (\ref{eq:pro1a}) for several values of $n$.}
\label{fig:onset}
\end{figure}

In Fig.~\ref{fig:onset} we show the results of our stability analysis. Our solutions all become unstable before we reach $a_{\max}$.  We plot the amplitude for the onset of the instability for  a variety of modes, $5 \leq m\leq 18$,  and for several profiles,  $2\leq n\leq 8$. In all cases,  the onset occurs for $a > a_{\ergo}$ as expected.  Independent of profile, the onset of the instability monotonically decreases with $m$, and appears to approach $a_{ergo}$ as $m\to \infty$. This is similar to the results found for Kerr AdS. It supports the idea that  an ergoregion is needed to have an instability, and the shortest modes become unstable first.
\subsubsection{Scalar condensate}

We now ask if there is a possible stationary endpoint for this instability. In the analogous problem involving gravity coupled to a Maxwell field, it was found that if one adds a charged scalar field,
the solutions also become unstable before $a_{\max}$. However, it was shown that there is a stationary solution with nonzero scalar field for all amplitudes \cite{Crisford:2017gsb}, so there is a natural endpoint to the instability. We now check if the same is true for our vacuum black holes coupled to a scalar field.  

It turns out to be convenient to work with a complex massless scalar field so we consider the Einstein-scalar action
\begin{equation}
S=\frac{1}{16\pi G_4}\int_{\mathcal{M}}\mathrm{d}^4x\,\sqrt{-g}\left(R+\frac{6}{L^2}-2\nabla_a \Phi\,\nabla^a \Phi^*\right)\,,
\end{equation}
where $^{*}$ denotes complex conjugation. We again use the Einstein-DeTurck equation, but including a scalar field. That is to say, we solve
\begin{subequations}
\begin{align}
&R_{ab}+\frac{3}{L^2}g_{ab}-\nabla_{(a}\xi_{b)}=\nabla_a \Phi\nabla_b \Phi^*+\nabla_b \Phi\nabla_a \Phi^*\,,
\\
&\Box \Phi =0\,.
\end{align}
\end{subequations}
To ease our numerical calculations, we consider a scalar field with a definite quantum number $m$:
\begin{equation}
\Phi = e^{i\,m\,\phi}\widehat{\Phi}\quad\text{and}\quad \Phi^* = e^{-i\,m\,\phi}\widehat{\Phi}\,,
\end{equation}
with $\widehat{\Phi}$ being real. The matter sector breaks axisymmetry, but the metric does not since the stress energy tensor only involves $\Phi$ in the combination $\nabla_{(a} \Phi\nabla_{b)} \Phi^*$. This is similar in spirit to the black holes with a single Killing field of \cite{Dias:2011at,Herdeiro:2014goa} and holographic Q-lattices of \cite{Donos:2013eha}. As before, we choose an \emph{ansatz} for our scalar field of the form:
\be
\widehat{\Phi}=\left(1-\tilde{x}^2\right)^3 \tilde{x}^{\left| m\right| } \left(2-\tilde{x}^2\right)^{\frac{\left| m\right| }{2}} \tilde{y}^{\left| m\right| +3} \left(2-\tilde{y}^2\right)^{\frac{\left| m\right| +3}{2}}q_7\,,
\ee
while our metric \emph{ansatz} remains as in Eq.~(\ref{eq:ansatz}).

\begin{figure}
\centering
    \includegraphics[width=0.45\textwidth]{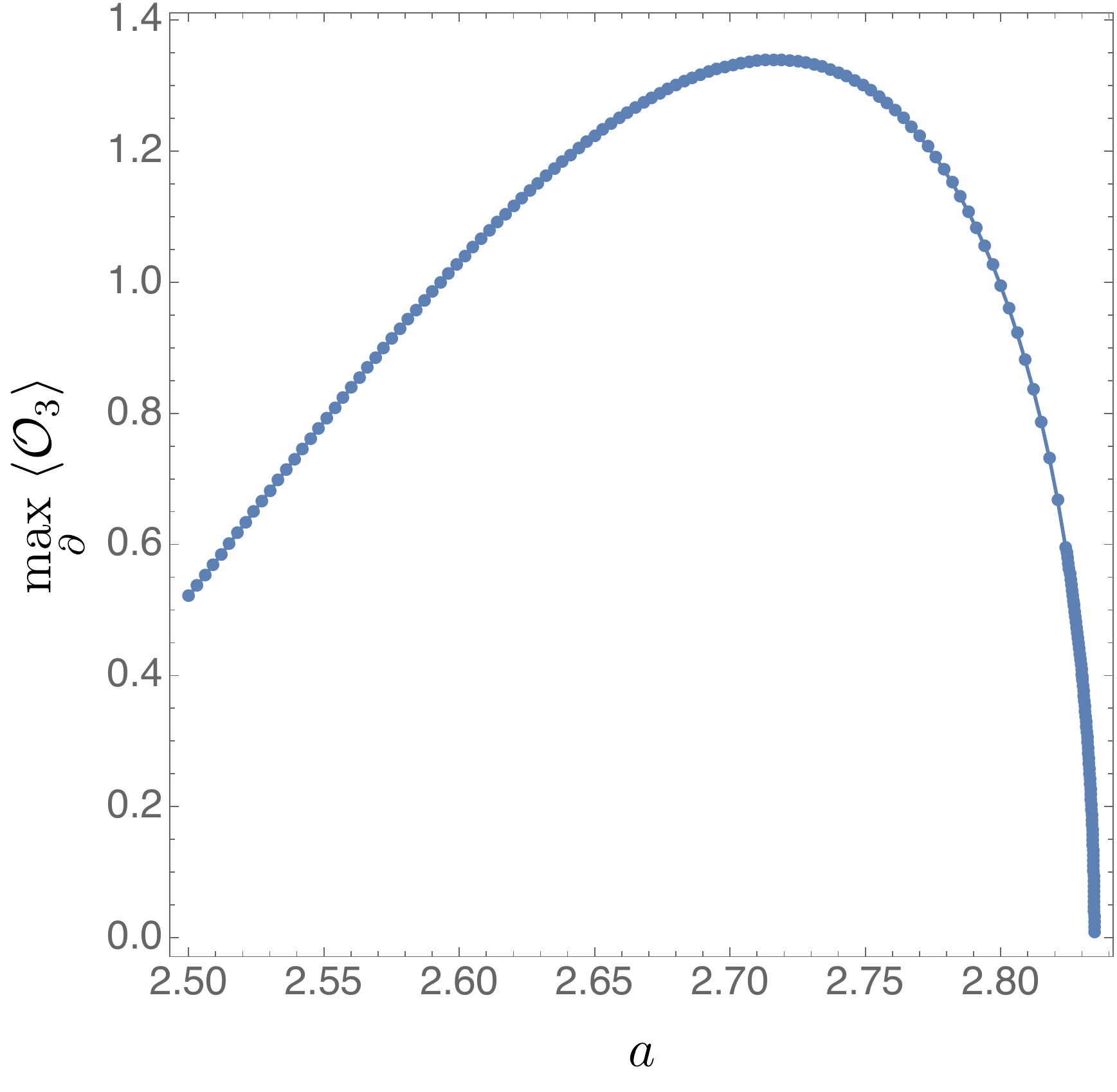}
  \caption{Solutions with a scalar condensate only exist for amplitudes {\it less} than the onset. The data shown here gives the maximum value of the condensate along the boundary for the $n=2$ profile and $m=6$ mode. The curve terminates at the left when the solution becomes singular. }
\label{fig:condensate}
\end{figure}

Our results are rather surprising and very different from the electromagnetic case. Stationary solutions with nonzero scalar field indeed branch off  from the  onset of the instability. However,  they now extend towards {\it smaller} values of $a$.  Eventually, these solutions become singular and terminate. This is depicted in Fig.~\ref{fig:condensate} for $n=2$ and $m=6$. To judge the size of the scalar field, we use the maximum of the expectation value of the dual scalar operator over the boundary, which is essentially the coefficient of the leading term in $\Phi$ as one approaches the boundary.  This is perhaps similar to the results found in \cite{Dias:2015rxy}, where the black resonators and black holes with a single Killing field of \cite{Dias:2011at} extend to smaller values of the angular velocity. Since our vacuum solutions are stable for these values of the amplitude, it is likely that these new solutions with nonzero scalar field are unstable\footnote{Any such solution will be unstable to higher $m$ perturbations. We are saying here that even for fixed $m$, these solutions are likely to be unstable.}. More importantly, there are no stationary configurations  for the larger amplitude solutions to settle down to, even including the scalar field. 

\subsection{\label{sec:blacknon}Black Holes}
We now extend our results to nonzero temperature, to see what happens if we start with a black hole rather than the vacuum.  The boundary metric will again be given by \eqref{eq:bndmetric}, and the differential rotation
 $\omega(r)$, will again take the form \eqref{eq:pro1a}.

\subsubsection{\label{eq:holeplanar}Metric \emph{ansatz}}
We will start by describing our choice of reference metric. First, we take the usual planar black hole written in the familiar Schwarzschild coordinates $(r,Z)$
\be
\mathrm{d}s^2=\frac{L^2}{Z^2}\left[-\left(1-\frac{Z^3}{Z_+^3}\right)\mathrm{d}t^2+\frac{\mathrm{d}Z^2}{\displaystyle1-\frac{Z^3}{Z_+^3}}+\mathrm{d}r^2+r^2\mathrm{d}\phi^2\right]\,,
\label{eq:planar}
\ee
where the horizon is the null hypersurface $Z=Z_+$ with the associated Hawking temperature
$T = {3}/{4\pi\,Z_+}$.
According to the gauge/gravity duality, the Hawking temperature will be identified with the field theory temperature $T$ \cite{Witten:1998qj}.

We now introduce new coordinates
\be
r=\frac{x\sqrt{2-x^2}}{1-x^2}\qquad \text{and} \qquad Z = Z_+(1-y^2)\,,
\ee
in terms of which the line element (\ref{eq:planar}) can be written as
\be
\mathrm{d}s^2=\frac{L^2}{(1-y^2)^2}\left\{-G(y)\,y_+^2\,y^2\mathrm{d}t^2+\frac{4\mathrm{d}y^2}{G(y)}+y_+^2\left[\frac{4\mathrm{d}x^2}{(2-x^2)(1-x^2)^4}+\frac{x^2(2-x^2)}{(1-x^2)^2}\mathrm{d}\phi^2\right]\right\}\,,
\ee
where $G(y)=3-3y^2+y^4$ and $y_+\equiv1/Z_+$. In terms of these new coordinates, the profile (\ref{eq:pro1a}) reduces to
\be
\omega(r)=  \frac{a}{(1+r^2)^{n/2}} \equiv g(x)=a\,(1-x^2)^n\,.
\label{eq:prox}
\ee

We now propose the following \emph{ansatz} for our metric
\begin{multline}
\mathrm{d}s^2=\frac{L^2}{(1-y^2)^2}\Bigg\{-G(y)\,y_+^2\,y^2\,q_1\mathrm{d}t^2+\frac{4\,q_2}{G(y)}\left[\mathrm{d}y+\frac{q_3\dd x}{(1-x^2)^2}\right]^2+\\
y_+^2\left[\frac{4\,q_4\mathrm{d}x^2}{(2-x^2)(1-x^2)^4}+\frac{x^2(2-x^2)}{(1-x^2)^2}\,q_5\left(\mathrm{d}\phi-y^2\,q_6\,\dd t\right)^2\right]\Bigg\}\,.
\label{eq:angen}
\end{multline}
For the reference metric we take Eq.~(\ref{eq:angen}) with $q_1=q_2=q_3=q_5=1$, $q_3=0$ and $q_6=g(x)$. 

We now discuss the issue of boundary conditions. Infinitely far away from the fixed points of $\partial_\phi$, \emph{i.e.} at $x=1$, we demand $q_1=q_2=q_3=q_5=1$ and $q_3=q_6=0$, which is consistent with our choice of profile (\ref{eq:prox}). At the centre, located at $x=0$, regularity demands
\be
\partial_x q_1=\partial_x q_2=\partial_x q_4=\partial_x q_5=\partial_x q_6=q_3=0\,.
\ee

At the conformal boundary, located at $y=1$, we choose the line element (\ref{eq:angen}) to approach the reference metric, \emph{i.e.}
\be
q_1=q_2=q_3=q_5=1\,,\qquad q_3=0\,,\qquad\text{and} \qquad q_6=g(x)\,.
\ee
Finally, at the horizon, located at $y=0$, regularity in ingoing Eddington-Finkelstein coordinates imposes
\be
\partial_y q_1=\partial_y q_2=\partial_y q_4=\partial_y q_5=\partial_y q_6=q_3=0\qquad\text{and}\qquad q_1=q_2\,,
\ee
with the later condition fixing the black hole temperature to be
\be
T = \frac{3}{4\pi} y_+\,.
\label{eq:hawking}
\ee

Lastly, we discuss how to extract the holographic stress energy tensor, following \cite{deHaro:2000vlm}. First, we solve (\ref{eq:einsteindeturck}) in a series expansion around conformal boundary $y=1$. This is done via the rather intricate expansion
\be
q_i = \sum_{j=0}^{+\infty}q_i^{(j)}(x)(1-y)^j+ (1-y)^{(3+\sqrt{33})/2}\sum_{j=0}^{+\infty}\widehat{q}_i^{(j)}(x)(1-y)^j+(1-y)^4\log(1-y)\sum_{j=0}^{+\infty}\tilde{q}_i^{(j)}(x)(1-y)^j\,.
\ee
The nonanalytic terms, which will affect the convergence of our numerical method, were first uncovered in \cite{Santos:2012he} and \cite{Donos:2014yya}.

Once the expansion is sorted out, the idea is to then change from our coordinates $(x,y)$ to Fefferman-Graham coordinates via a new asymptotic expansion
\begin{subequations}
\begin{align}
&x =\sqrt{1-\frac{1}{\sqrt{1+r^2}}}+\alpha_1(r)\,z+\alpha_2(r)\,z^2+\alpha_3(r)\,z^3+\alpha_4(r)\,z^4+\smallO(z^4)
\\
&y=1+\beta_1(r)\,z+\beta_2(r)\,z^2+\beta_3(r)\,z^3+\beta_4(r)\,z^4+\smallO(z^4)
\end{align}
\end{subequations}
and determine the coefficients $\alpha_i$ and $\beta_i$ by requiring the line element (\ref{eq:angen}) to be in the Fefferman-Graham form, \emph{i.e.}
\be
\dd s^2=\frac{L^2}{z^2}\left[\dd z^2 + \dd s_0^2+\dd s_2^2\,z^2+\dd s_3^2\,z^3+\smallO(z^4)\right]\,,
\ee
where $\dd s_i^2$, with $i\in\{0,2,3\}$, only has components along the boundary directions. For instance, we find
\be
\alpha_1=0\qquad\text{and}\qquad \beta_1=-\frac{y_+}{2}\,.
\ee

After recasting the metric is in this form, the holographic stress energy tensor is recovered via \cite{deHaro:2000vlm}
\be
\langle T_{\mu\nu}\rangle \dd x^\mu \dd x^\nu = \frac{3}{16\,\pi} \dd s_3^2\,.
\ee
\subsubsection{Results}
\begin{figure}
\centering
    \includegraphics[width=0.45\textwidth]{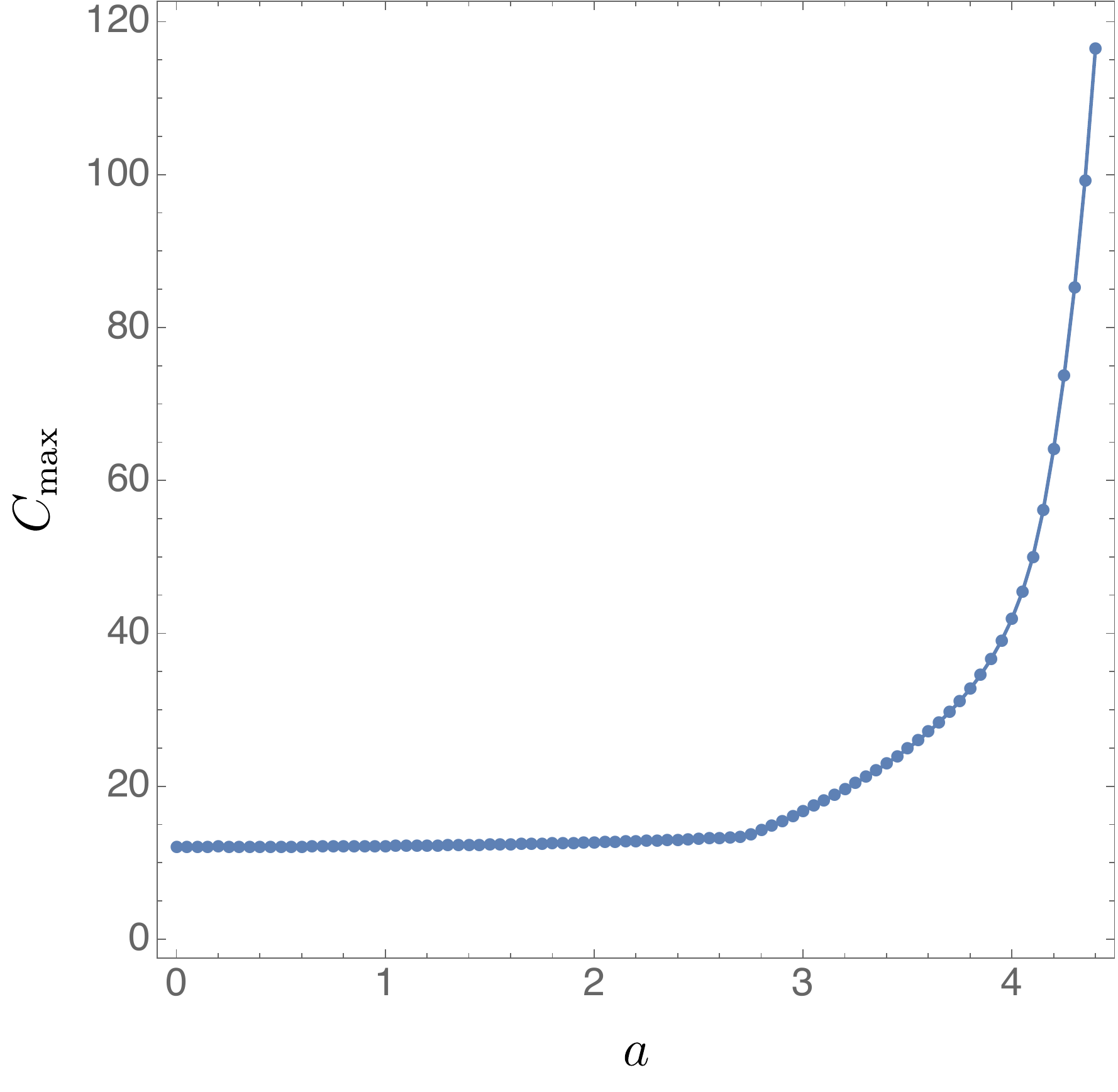}
  \caption{$C_{\max}$ as a function of $a$, computed for $n=4$ and $T=.9/4\pi$.}
\label{fig:cmaxt}
\end{figure}

\begin{figure}
\centering
    \includegraphics[width=0.45\textwidth]{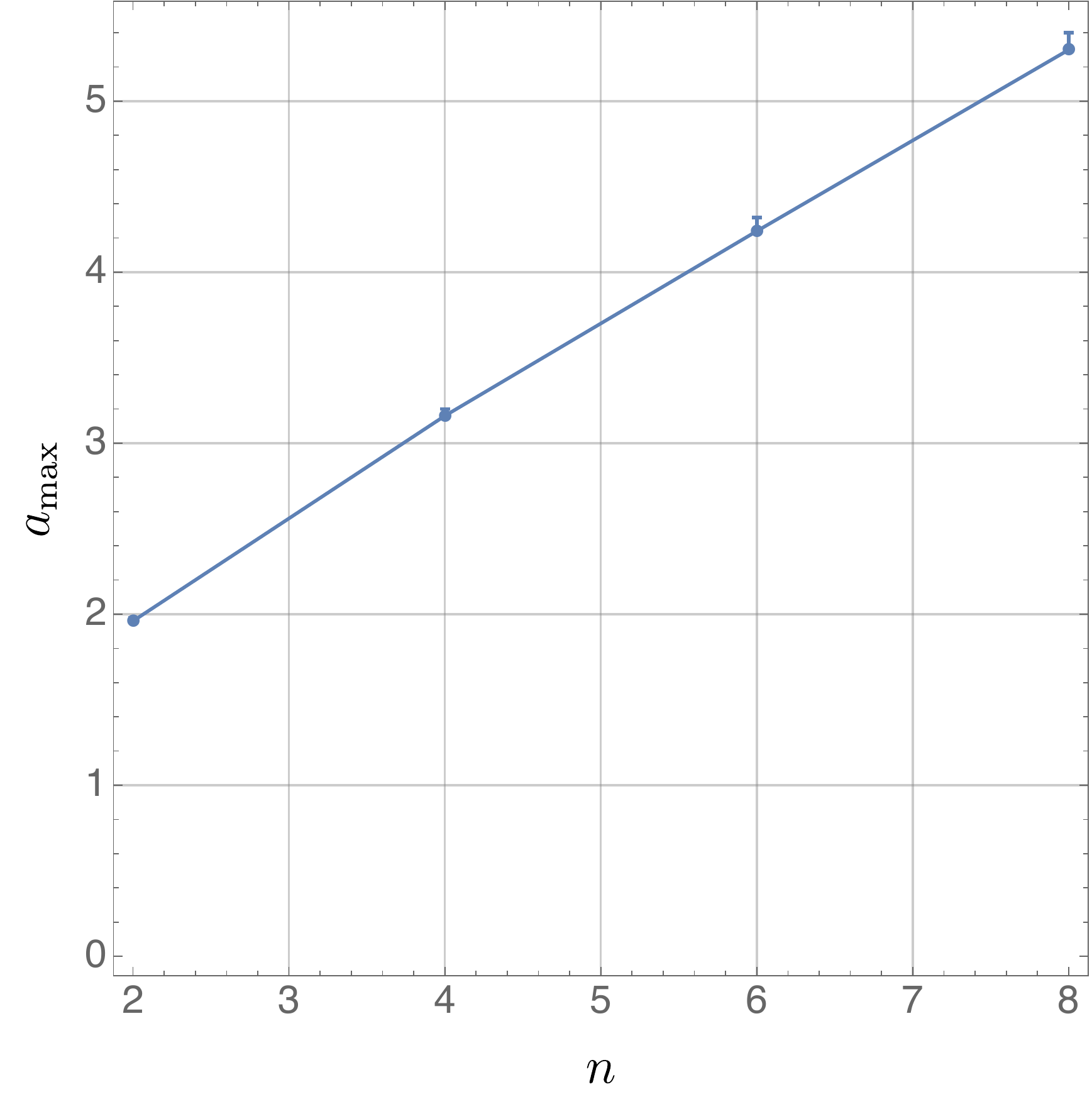}
  \caption{$a_{\max}$ as a function of $n$, computed for $T = 3/4\pi$.}
\label{fig:amaxseveraln}
\end{figure}

We have followed the same procedure as before, increasing the amplitude $a$ for fixed temperature $T$, and for several different profiles: $n=2,4,6,8$. In all  cases, we again find that there is a maximum amplitude we can attain before our solution becomes singular\footnote{This was not true in the analogous electromagnetic problem, where static solutions were found with $T>0$ for any amplitude of the chemical potential \cite{Horowitz:2016ezu}.}. As a typical example,  in Fig.~\ref{fig:cmaxt} we plot the maximum value of the square of the Weyl tensor, $C_{\max}$  \eqref{eq:cmax} for the case $n=4$ and  $T = .9/4\pi$.   The apparent kink in Fig.~\ref{fig:cmaxt} results from the crossing of two local maxima, analogous to the
absolute maximum going through a first order phase transition.
As in the $T=0$ case,   $a_{\max}$ increases with $n$. This is illustrated in Fig.~\ref{fig:amaxseveraln},  where we plot
 $a_{\max}$ at fixed $T = 3/4\pi$,  for several values of $n$. For all profiles, $a_{\max} > a_{\ergo}$.

\begin{figure}
\centering
    \includegraphics[width=0.45\textwidth]{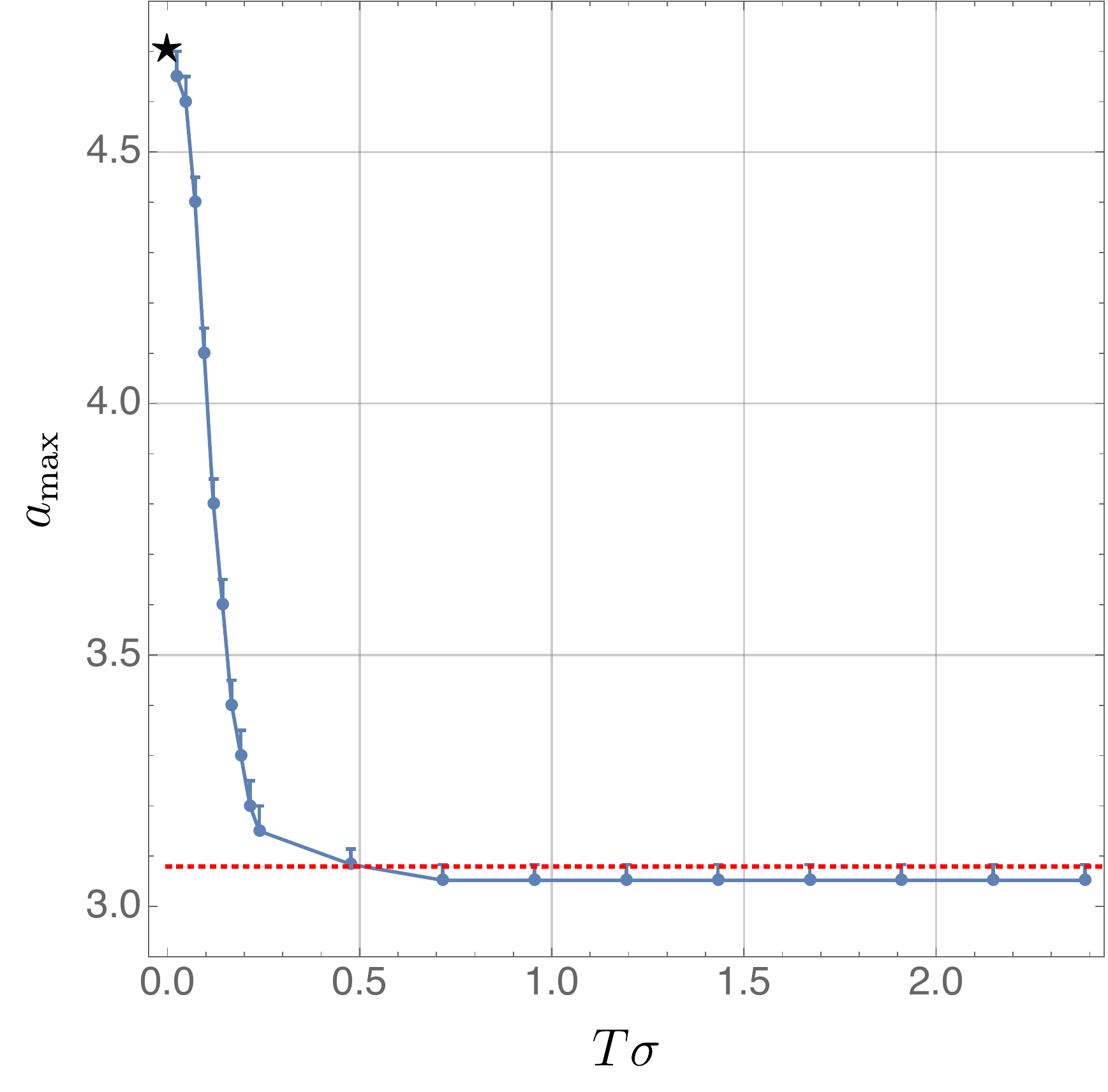}
  \caption{$a_{\max}$ as a function of $T$, computed for $n=4$. The black star is the $T=0$ result and the dotted red line is $a_{\ergo}$.}
\label{fig:amaxt0}
\end{figure}

Next,  we examine how $a_{\max}$ changes when we turn up the temperature. We find that $a_{\max}$ decreases rapidly from its $T=0$ value and settles down to  $a_{\ergo}$. To illustrate this,  in Fig.~\ref{fig:amaxt0} we plot $a_{\max}$ as a function of $T$, for fixed $n=4$. The black dot is the $T=0$ value, the dotted red line is $a_{\ergo}$, and the blue dots are the numerical values  of $a_{\max}$ we extracted at finite temperature.

Despite appearances in Fig.~\ref{fig:amaxt0}, we do not expect $a_{\max} = a_{\ergo}$ at finite $T$, but only to approach it from above as $T\to\infty$. This is because one can construct stationary black holes with large $T$ for all $a < a_{\ergo}$ using holography, and in particular, the fluid/gravity correspondence \cite{Bhattacharyya:2008jc,Bhattacharyya:2008xc,Hubeny:2011hd}. When the scale of curvature is much larger than the thermal wavelength, one expects the dual field theory will be well described by a fluid.  In this case, the fluid/gravity correspondence constructs a bulk solution by associating a piece of a boosted planar black hole to  boundary regions that are smaller than the curvature scale but larger than the thermal wavelength, and suitably patching them together.  For  $a<a_{\ergo}$ one can indeed construct  stationary bulk black holes using this procedure which agree very well with our numerical solutions. However one cannot obtain stationary solutions this way when there is an ergoregion on the boundary, since the Killing field becomes spacelike and can no longer define a local rest frame for the fluid. One can presumably construct nonstationary black holes by picking a slowly varying unit timelike vector on the boundary to use as the fluid four-velocity.

\begin{figure}
\centering
    \includegraphics[width=\textwidth]{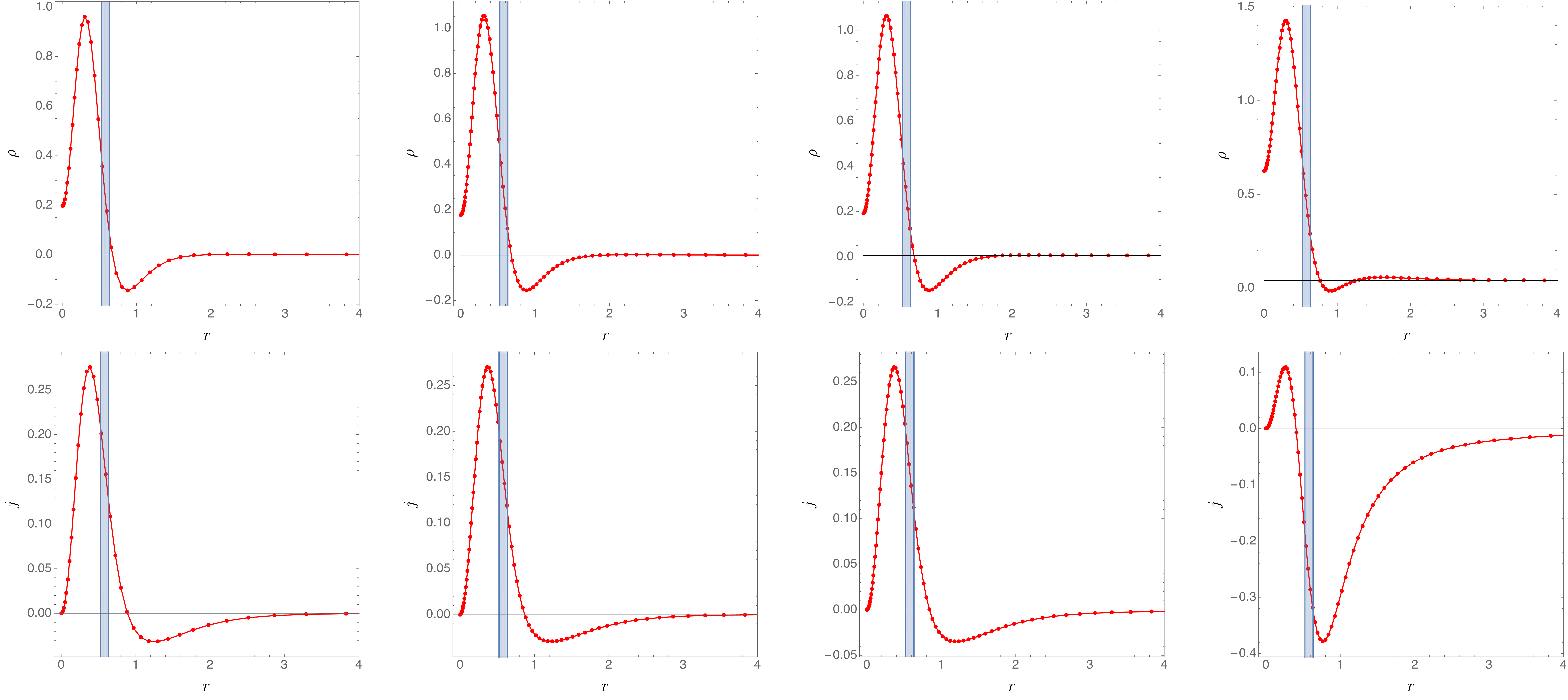}
  \caption{The holographic energy density (top row) and holographic momentum density (bottom row) computed for $n=4$ and fixed $a=3.1$ at four distinct temperatures. The shaded region denotes the ergoregion.  From left to right we have $4\pi T/3=0,0.1,0.5,1$. 
  The horizontal solid black line in the holographic energy density plots indicate the value that these quantities take for a planar Schwarzschild black brane (\ref{eq:planar}).}
\label{fig:holot}
\end{figure}

Finally, we have studied how the energy density $\rho$ and  angular momentum density $j$ \eqref{eq:thermo} depend on the temperature. This is shown in Fig.~\ref{fig:holot} for $4\pi T/3 =0.0,0.1,0.5,1.0$ at fixed $n=4$ and $a=3.1$, which is close to $a_{\max}$.  The shaded regions correspond to the location of the boundary ergoregion.  
The horizontal solid black line in the holographic energy density plots indicate the value that these quantities take for a planar Schwarzschild black hole (\ref{eq:planar}). Note that $j$ vanishes identically for the planar Schwarzschild black hole (\ref{eq:planar}).  The figure shows  that $\rho$ and  $j$ differ significantly from their Schwarzschild values in the vicinity of the ergoregion.   The curves corresponding to $T=0$ were taken from the analysis of the previous section (corresponding to the two plots on the left column of Fig.~\ref{fig:holot}). 
The total angular momentum $J$ is now nonzero and grows with $T$. Unlike the total energy $E$, $J$ appears to diverge as $a\to a_{\max}$.

\section{Compact solutions}

In this section we consider a different class of solutions where the boundary has topology $\mathbb{T}^2\times \mathbb{R}$ and we add rotation around the circles.  We will mostly focus on the case where we add rotation around one circle, so  the boundary metric takes the form 
\be\label{eq:rotcircle}
\mathrm{d}s_\partial^2=-\dd t^2+{\mathrm{d}X^2}+\left[{\mathrm{d}W}-\omega(X)\mathrm{d}t\right]^2\,,
\ee
where $X$ and $W$ are both periodic with periods $\ell_X = 2\pi/k_X$ and $\ell_W=2\pi/k_W$ respectively, and
\be
\omega(X)=a\,\cos k_X X\,.
\label{eq:protor}
\ee
  Note that if $a>1$, there is a boundary ergoregion, whereas if $a<1$ there is none, so $a_{\ergo}= 1$. The case $a=1$  corresponds to the situation when we have an evanescent ergosurface.
Unlike  the non-compact case, $a$ now always has conformal dimension $0$, so it corresponds to a marginal deformation of the boundary metric.

 At the end of this section, we will briefly comment on what happens if we add rotation to both circles and the boundary metric takes the form:
\be\label{eq:rotcircle2}
\mathrm{d}s_\partial^2=-\dd t^2+\left[\mathrm{d}X - \tilde\omega(W) \mathrm{d}t\right]^2 +\left[{\mathrm{d}W}-\omega(X)\mathrm{d}t\right]^2\,,
\ee
with
\be
\tilde \omega(W)=\tilde a\,\cos k_W W\,.
\label{eq:protorW}
\ee

\subsection{Zero-temperature solutions}
We first discuss  zero-temperature solutions. Even though our ansatz for the boundary metric \eqref{eq:rotcircle} has three parameters $a,k_X,k_W$, there is only a one parameter family of inequivalent solutions labelled by $a$. This is because we can use scale invariance to set $k_X = 1$, and since our boundary metric  is independent of $W$, our solution will be independent of $W$ and $\ell_W$ will only appear as an overall factor.

Since our boundary metrics  are compact, we have a couple of possibilities for the IR behaviour of our solutions. Namely, one can have a solitonic solution, with no horizons, where a spatially compact direction smoothly caps off spacetime \cite{Horowitz:1998ha}, or we can try to compactify the Poincar\'e horizon.   If one starts with the Poincar\'e patch of AdS and makes the spacelike directions of the Minkowski slices compact by periodically identifying them, the horizon develops a conical singularity and is no longer smooth.  This is because the translational symmetries have a fixed point there. 

In a canonical ensemble,   the solution that is likely to dominate at $T=0$  is the solitonic one, since this is true without the rotation. However, we are interested in studying these solutions from a microcanonical perspective, since that is appropriate when evolving at fixed energy. Furthermore,  we are interested in the $T\to 0$ limit of the black holes we will construct in part {\bf B}, so we will focus on the solutions with a compactified Poincar\'e horizon. We leave the construction of the solitonic solutions to a future endeavour.

\subsubsection{Metric ansatz }

Again, we use the DeTurck method which we outlined in section \ref{sec:turck} to construct solutions.  We start with pure AdS written in familiar Fefferman-Graham coordinates
\be
\mathrm{d}s^2=\frac{L^2}{Z^2}(-\dd t^2+\dd X^2+\dd W^2+\dd Z^2)\,,
\label{eq:almostfinal}
\ee
where again $X$ and $W$ are periodic coordinates with period $\ell_X\equiv2\pi/k_X$ and $\ell_W\equiv2\pi/k_W$. We now introduce coordinates
\begin{equation}
Z=\frac{y\sqrt{2-y^2}}{1-y^2}\,,\qquad X = \frac{x}{k_X}\,,\qquad \text{and}\qquad W=\frac{w}{k_W}\,,
\end{equation}
which brings Eq.~(\ref{eq:almostfinal}) into the following form
\be
\mathrm{d}s^2=\frac{L^2}{y^2(2-y^2)}\left[(1-y^2)^2\left(-\dd t^2+\frac{\dd x^2}{k_X^2}+\frac{\dd w^2}{k_W^2}\right)+\frac{4\dd y^2}{(2-y^2)(1-y^2)^2}\right]\,.
\label{eq:almostfinal2}
\ee
The form of (\ref{eq:almostfinal2}) is ideal to introduce the DeTurck trick, since $y=1$ marks the Poincar\'e horizon and $y=0$ the location of the conformal boundary while the remaining two boundary coordinates have period $2\pi$. In this section we will only study rotation profiles along the $W$ direction, which explicitly depend on $X$. Our line element for the DeTurck method reads
\begin{multline}
\mathrm{d}s^2=\frac{L^2}{y^2(2-y^2)}\Bigg\{(1-y^2)^2\left[-q_1\,\dd t^2+q_4\,\left(\frac{\dd x}{k_X}+q_3\dd y\right)^2+q_5\left(\frac{\dd w}{k_W}-q_6(1 - y^2)^2 \dd t\right)^2\right]\\+\frac{4\,q_2\dd y^2}{(2-y^2)(1-y^2)^2}\Bigg\}\,,
\label{eq:almostfinal3}
\end{multline}
which is invariant under general reparametrizations of $(x,y)$. For the reference metric we will take
\be
q_1=q_2=q_4=q_5=1\,,\qquad q_3=0\,,\qquad \text{and}\qquad q_6 = a\,\cos x\,.
\label{eq:referencezerocompact}
\ee

The boundary conditions at the IR of the theory, that is to say at the horizon located at $y=1$, are simply
\be
\partial_y q_1=\partial_y q_4=\partial_y q_5=0\,,\qquad q_2=1\,\qquad\text{and}\qquad q_3=q_6=0,
\ee
while at the conformal boundary we demand our physical bulk spacetime metric to approach the reference metric (\ref{eq:referencezerocompact}). We shall see that the IR will depend on $a$, but in a trivial way. In particular, the IR will always be Poincar\'e, but $g_{tt}$, $g_{ww}$ and $g_{xx}$ will appear renormalised along the RG flow as we move from the UV to the IR. This is to be expected, since from the perspective of the UV theory, $a$ is a marginal deformation. The boundary conditions above are compatible with such IR behaviour. We should \emph{a posteriori} check that $q_1$, $q_4$ and $q_5$ are independent of $x$ when $y=1$, which will turn out to be the case for all the runs we have made.

\subsubsection{Results}
\begin{figure}
\centering
    \includegraphics[width=0.50\textwidth]{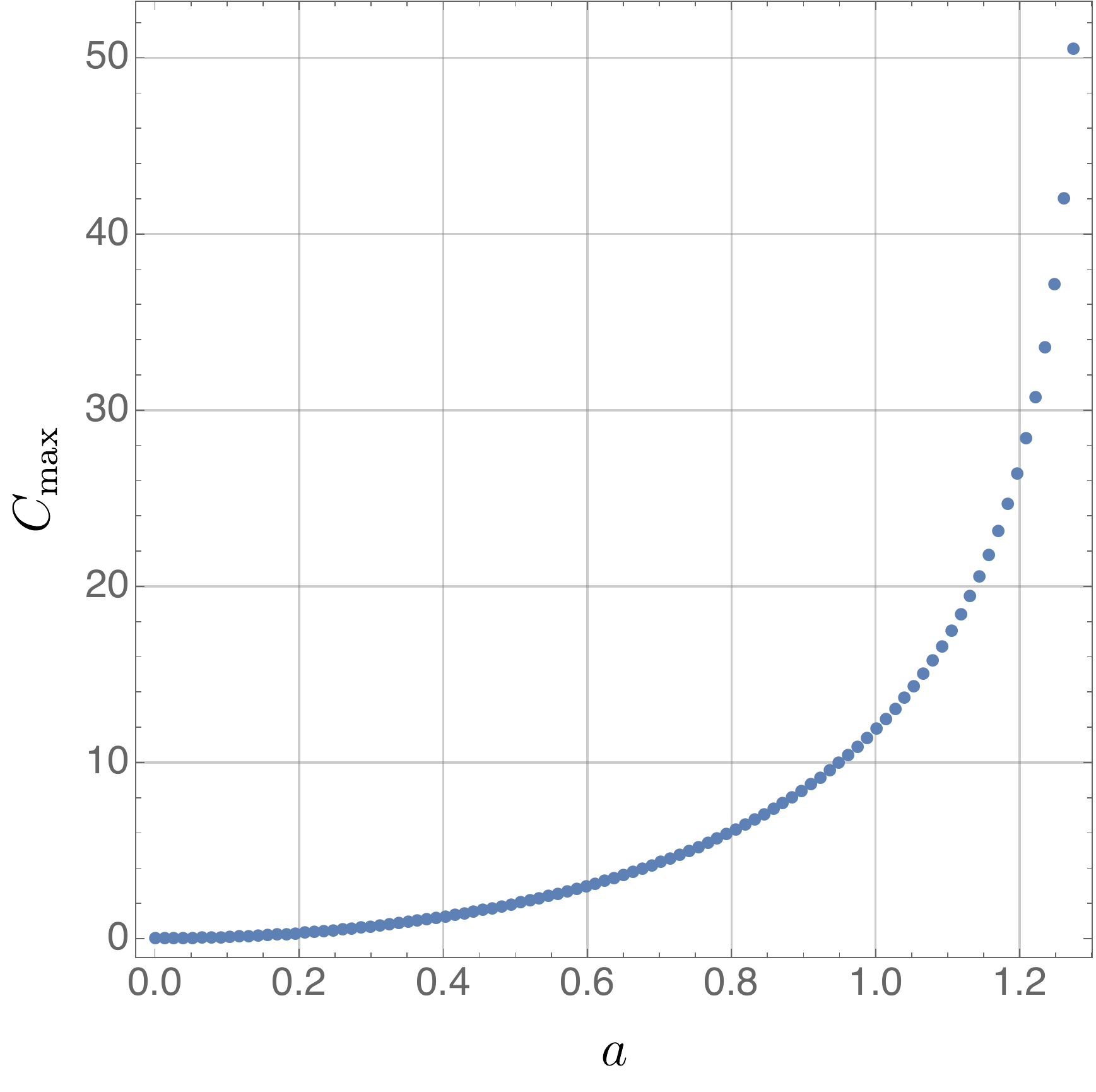}
  \caption{$C_{\max}$ as a function of $a$ for the $T=0$ compact case.}
\label{fig:cmaxzerotcompact}
\end{figure}

\begin{figure}
\centering
    \includegraphics[width=0.45\textwidth]{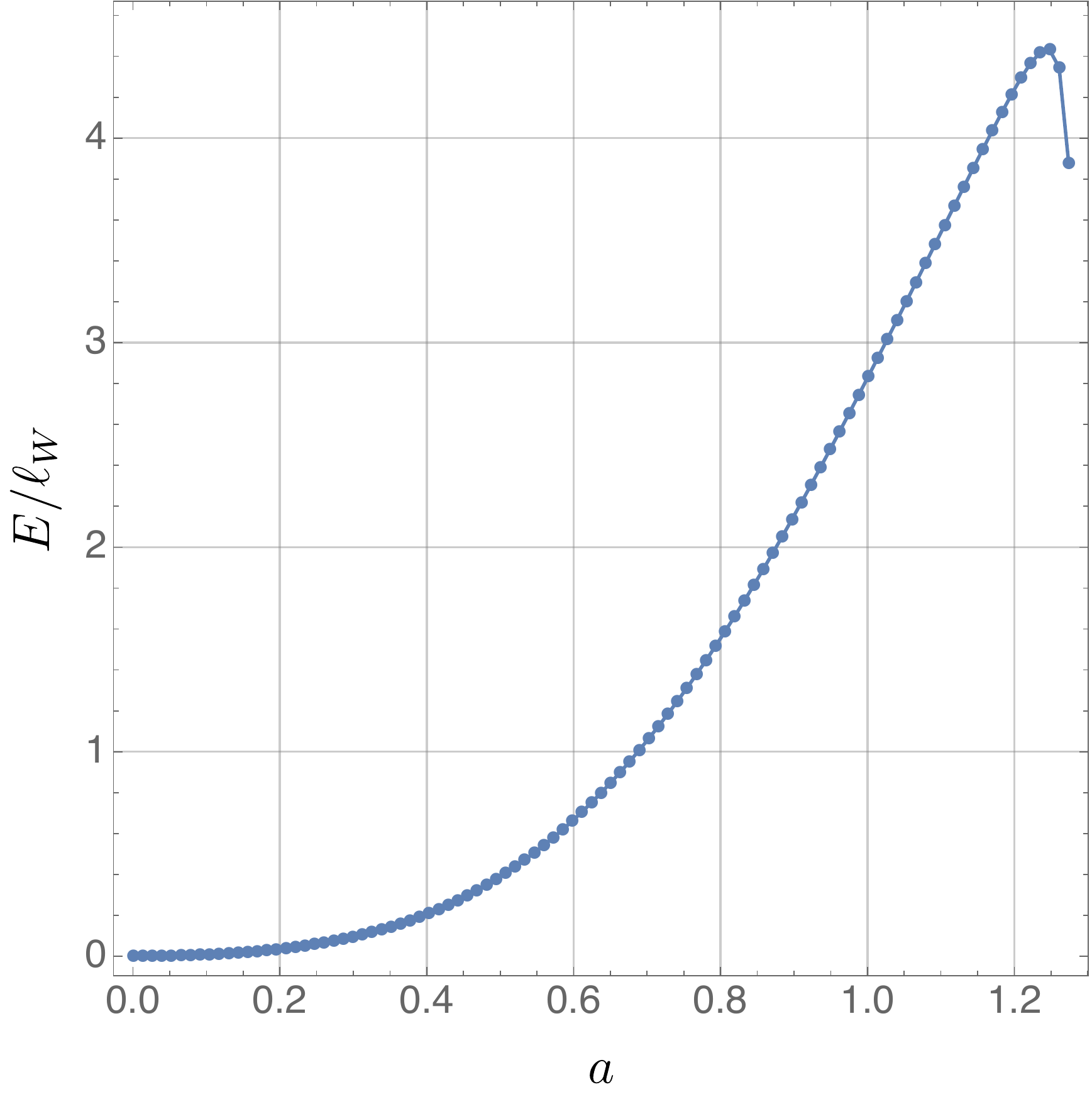}
  \caption{The energy density $E/\ell_W$ as a function of $a$, with the non-monotonic behaviour starting at around 
  $a\approx1.24422$.}
\label{fig:energytotalcompact}
\end{figure}

Just as in the noncompact case, there is a maximum amplitude $a_{\max}$ beyond which the solutions develop a curvature singularity. To find $a_{\max}$, we again monitor the maximum value of the square of the Weyl tensor, $C_{\max}$, as a function of $a$ and determine where it diverges. This is plotted in Fig.~{\ref{fig:cmaxzerotcompact}}. 
Like the non-compact zero-temperature solutions of section \ref{sec:resultsT0}, solutions exists even when $a>a_{\ergo}=1$. In fact we find $a_{\max}\approx1.28$.
We have computed other quantities such as $\rho$ and $j$, but they behave just like in the non-compact case, so we will not present them here. One of the quantities of interest that we can extract from these is the energy density $E/\ell_W$ as a function of $a$. This is presented in Fig.~\ref{fig:energytotalcompact}, where we again see $E$ increasing monotonically even past $a=a_{\ergo}$, but reaching a maximum value just before $a_{\max}$. Just like for the non-compact case, we have no current understanding of this behaviour.

\subsection{Black Holes}

We next discuss black hole solutions with boundary conditions \eqref{eq:rotcircle} or \eqref{eq:rotcircle2}. Since the temperature $T$ is a new dimensionful parameter, 
inequivalent solutions can now depend on the dimensionless ratios
$T/k_X$ and $T/k_W$. Increasing the wavenumbers at fixed temperature will have the same effect as decreasing the temperature. Combined with the dimensionless amplitude $a$, our moduli space is thus either two or three-dimensional depending on which  boundary condition we impose.

\subsubsection{Metric ansatz}
We will again use the DeTurck method. We will start by recalling the line element of a Schwarzschild black brane with toroidal spatial cross sections
\be
\mathrm{d}s^2=\frac{L^2}{Z^2}\left[-\left(1-\frac{Z^3}{Z_+^3}\right)\mathrm{d}t^2+\frac{\mathrm{d}Z^2}{\displaystyle1-\frac{Z^3}{Z_+^3}}+\mathrm{d}X^2+\mathrm{d}W^2\right]\,,
\label{eq:planartorus}
\ee
and we are interested in the case where both $X$ and $W$ are periodic with periods $\ell_X\equiv 2\pi/k_X$ and $\ell_W\equiv 2\pi/k_W$, respectively. Next, we change to new variables
\be
Z=Z_+(1-y^2)\,,\qquad X=\frac{x}{k_X}\qquad\text{and}\qquad W=\frac{x}{k_W}
\ee
in terms of which (\ref{eq:planartorus}) can be recast as
\be
\mathrm{d}s^2=\frac{L^2}{(1-y^2)^2}\left[-G(y)\,y_+^2\,y^2\,\mathrm{d}t^2+\frac{4\mathrm{d}y^2}{G(y)}+\frac{y_+^2}{k_X^2}\mathrm{d}x^2+\frac{y_+^2}{k_W^2}\mathrm{d}w^2\right]\,,
\label{eq:planartorus}
\ee
where $G(y)=3-3\,y^2+y^4$, $y_+\equiv 1/Z_+$ and $x$, $w$ are periodic coordinates with period $2\pi$. The horizon is the null hypersurface $y=0$, and has Hawking temperature
\begin{equation}
T = \frac{3\,y_+}{4\pi}\,.
\label{eq:hawkingto}
\end{equation}

We can now detail the \emph{ansatz} we used in the DeTurck method. We recall that this \emph{ansatz} should be compatible with diffeomorphism invariance in the $(x,y)$ directions. The line element reads
\be
\mathrm{d}s^2=\frac{L^2}{(1-y^2)^2}\Bigg\{-G(y)\,y_+^2\,y^2\,q_1\,\mathrm{d}t^2+\frac{4\,q_2\mathrm{d}y^2}{G(y)}+\frac{y_+^2}{k_X^2}q_4\left(\mathrm{d}x+q_3\mathrm{d}y\right)^2+y_+^2\,q_5\left(\frac{\mathrm{d}w}{k_W}-y^2\,q_6\,\dd t\right)^2\Bigg\}\,,
\ee
where all six functions $q_i$ are functions of $(x,y)$ only. For the reference metric in the DeTurck method, we will use the line element above with
\be
q_1=q_2=q_4=q_5=1\,,\quad q_3=0\,,\quad \text{and}\quad q_6=\omega(x)\,.
\label{eq:refto}
\ee

The boundary conditions are determined by requiring regularity across the event horizon, which demands
\be
\partial_y q_1=\partial_y q_2=\partial_y q_4=\partial_y q_5=\partial_y q_6=0\,,\qquad q_3=0\quad\text{and}\quad q_1=q_2\,.
\ee
Note that the last boundary condition ensures that the black hole temperature is given as in (\ref{eq:hawkingto}). At the boundary, we give Dirichlet boundary conditions and demand the metric to approach the reference metric (\ref{eq:refto})

Finally, we will also briefly discuss the case where we have boundary deformations in both the $x$ and the $w$ directions. This corresponds to a full three dimensional problem, where the black hole has a single Killing isometry corresponding to time translations $\partial/\partial t$. The most general line element compatible with such reduced symmetries reads
\begin{multline}
\dd s^2=\frac{L^2}{(1-y^2)^2}\Bigg[-G(y)\,y_+^2\,y^2\,Q_1\,\mathrm{d}t^2+\frac{4\,Q_2}{G(y)}\left(\mathrm{d}y+y^2Q_7 \mathrm{d}t\right)^2+y_+^2 Q_3\left(\frac{\mathrm{d}x}{k_X}-y^2Q_5\mathrm{d}t+Q_8\mathrm{d}y\right)^2\\
+y_+^2\,Q_4\left(\frac{\mathrm{d}w}{k_W}-y^2Q_6\,\dd t+Q_9\dd y+Q_{10}\dd x\right)^2\Bigg]
\end{multline}
where all ten functions $Q_i$ are functions of $(x,y,w)$. For the reference metric we now choose
\begin{align}
&Q_i=1, \qquad\text{for}\qquad i\in\{1,2,3,4\}\,,\nonumber
\\
&Q_i=0, \qquad\text{for}\qquad i\in\{7,8,9,10\}\,,\label{eq:refcrazy}
\\
&Q_5=a_x\,\cos w\quad\text{and}\quad Q_6=a_w\,\cos x\,.\nonumber
\end{align}

The boundary conditions at the horizon again following from requiring regularity across the event horizon
\begin{align}
&\partial_y Q_i=1, \qquad\text{for}\qquad i\in\{1,2,3,4,5,6\}\,,\nonumber
\\
&Q_i=0, \qquad\text{for}\qquad i\in\{7,8,9,10\}\,,
\\
&Q_1=Q_2\,,\nonumber
\end{align}
with the later fixing the temperature to be given by (\ref{eq:hawkingto}). Finally, at the conformal boundary, we demand the bulk physical metric to approach the reference metric (\ref{eq:refcrazy}).

\subsubsection{Results}
\begin{figure}
\centering
    \includegraphics[width=0.45\textwidth]{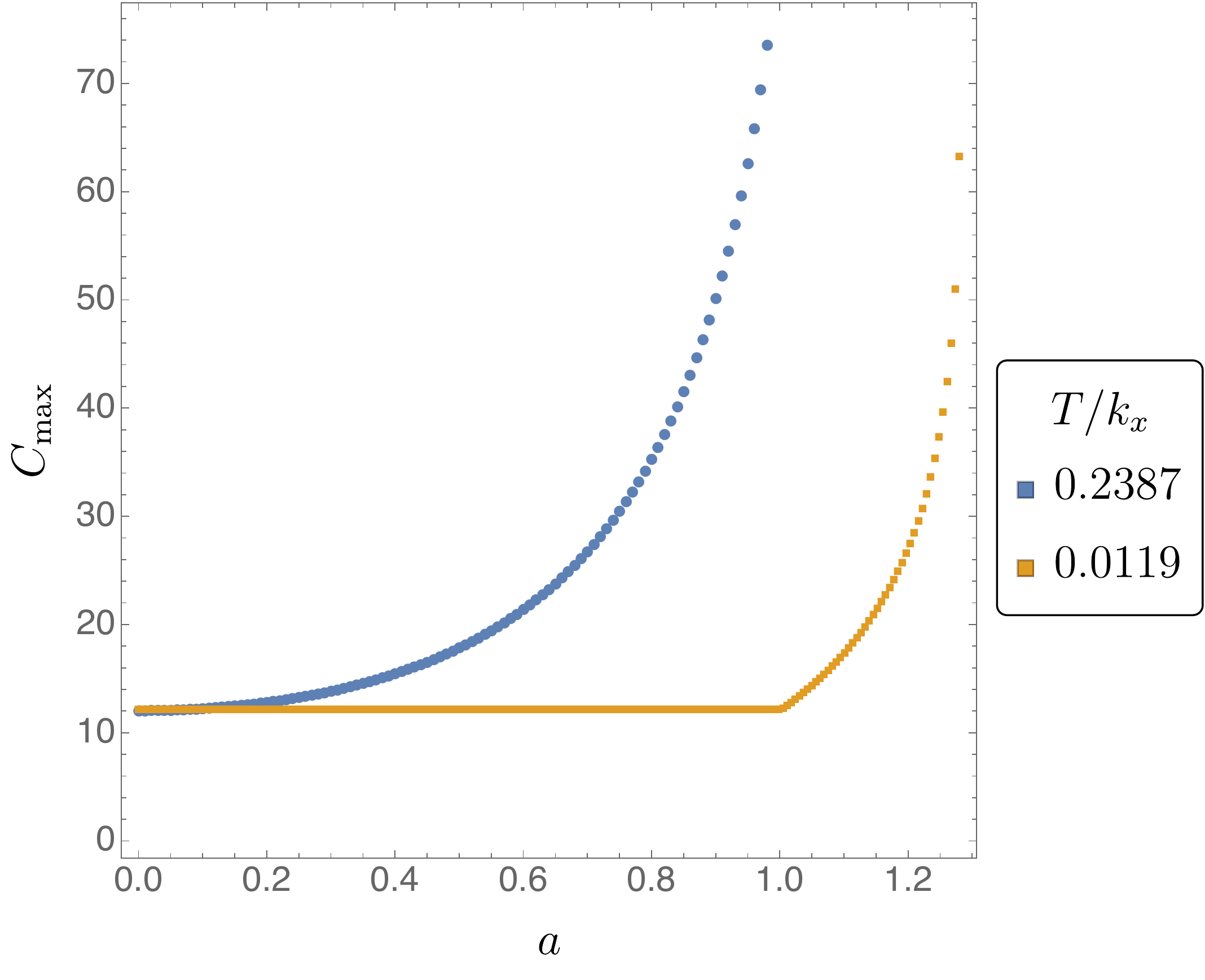}
  \caption{$C_{\max}$ as a function of $a$, depicted for $T/k_X=0.2387$ (top curve) and $T/k_X=0.0119$ (bottom curve). The kink in the bottom curve corresponds to the interchange of two local maxima.}
\label{fig:cmaxtoroidal}
\end{figure}

We start by adding rotation around one circle.
Just as in section \ref{sec:blacknon} we find that solutions exist only up to a maximum value $a_{\max}$, which strongly depends on the ratio $T/k_X$. We first fix $T/k_X$ and increase $a$ until the curvature, $C_{\max}$, appears to diverge. This is depicted in Fig.~\ref{fig:cmaxtoroidal} for $T/k_X=0.239$ (top curve) and $T/k_X=0.0119$ (bottom curve).

\begin{figure}
\centering
    \includegraphics[width=0.45\textwidth]{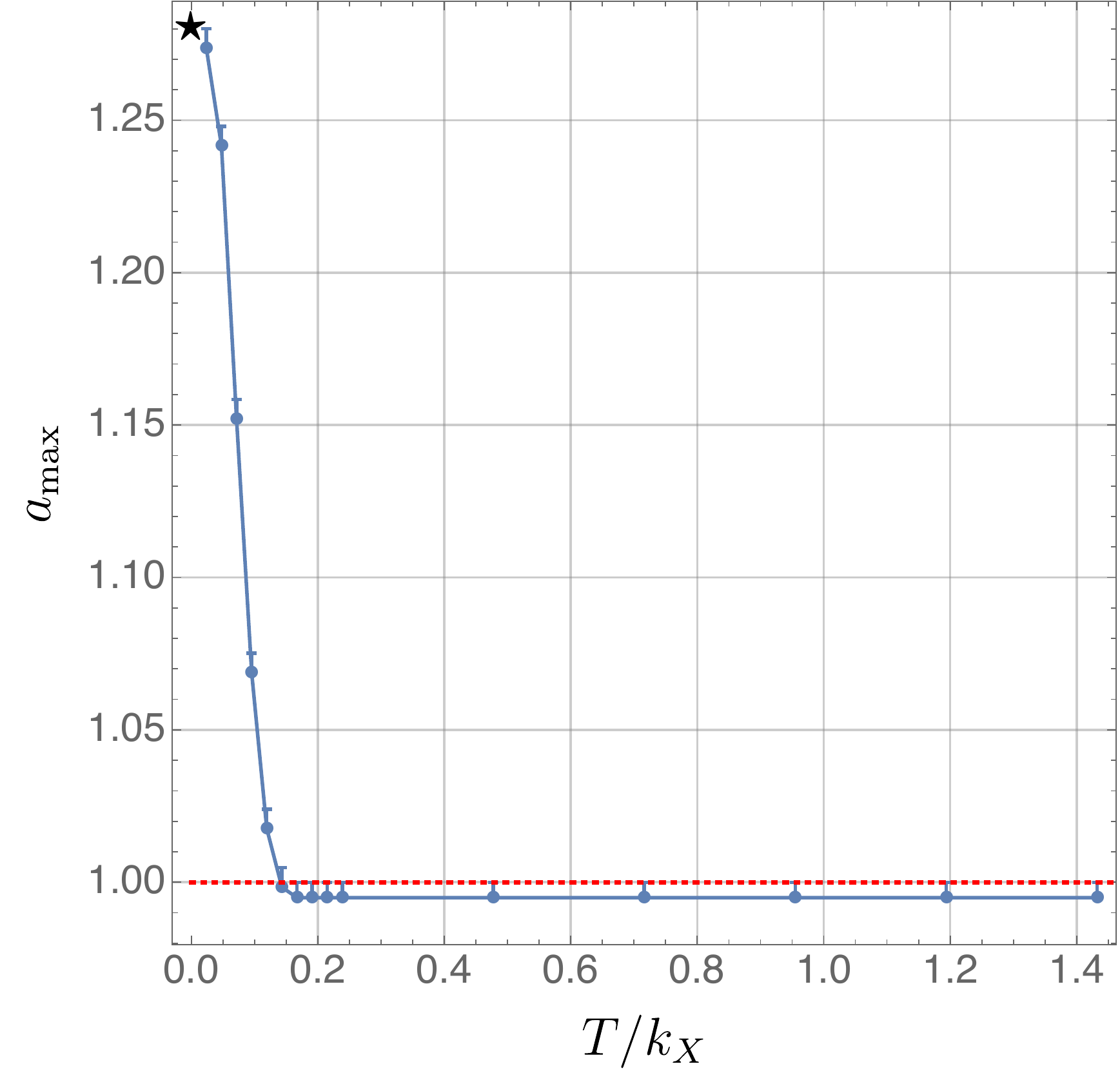}
  \caption{$a_{\max}$ as a function of $T/k_X$. The black star is the $T=0$ result obtained in the previous section and the dotted red line is $a=a_{\ergo}=1$.}
\label{fig:amaxt0c}
\end{figure}

In the next step, we investigate how $a_{\max}$ depends on $T/k_X$ by repeating the same calculation that leads to Fig.~\ref{fig:cmaxtoroidal} for many values of $T/k_X$. The results are plotted in Fig.~\ref{fig:amaxt0c}. Again we see that $a_{\max}$ decreases rapidly from the $T=0$ result computed in the previous section  to $a_{\ergo} =1$ in the fluid limit (corresponding to the high-temperature regime).

\begin{figure}
\centering
    \includegraphics[width=0.9\textwidth]{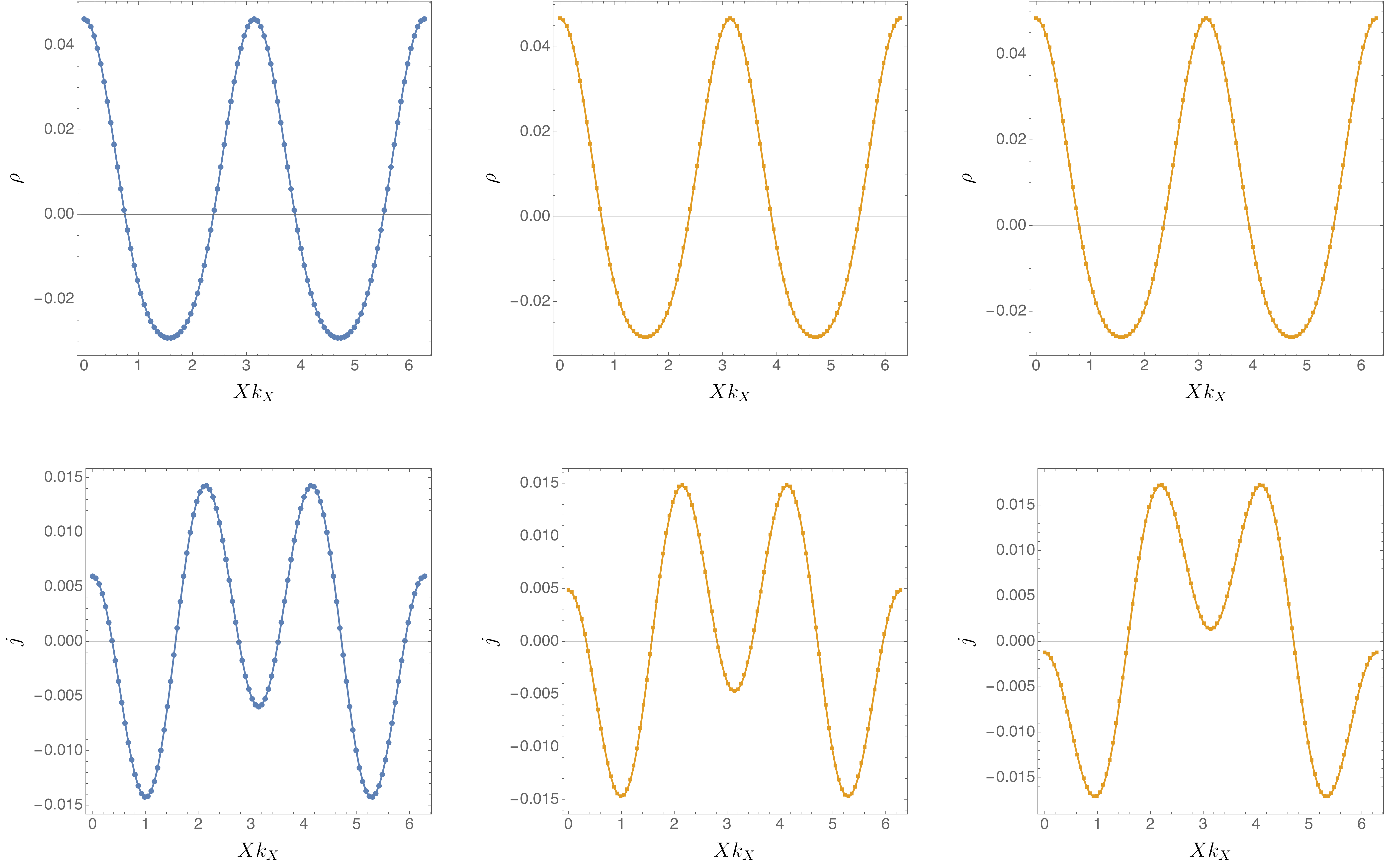}
  \caption{The holographic energy density (top row) and holographic momentum density (bottom row) computed at fixed $a=1.1008$ at three distinct temperatures. From left to right we have $T/k_X=0.0239,0.0477,0.0716$.}
\label{fig:holont}
\end{figure}

We also studied how $\rho$ and $j$ depend on $T/k_X$ at fixed $a=1.1008$, which can be seen in Fig.~\ref{fig:holont} for $T/k_X=0.0239,0.0477,0.0716$. Since $a >1$ there is an ergoregion on the boundary, but for these low temperatures, $\rho$ and $j$ change only modestly with $T$.

\begin{figure}
\centering
    \includegraphics[width=0.9\textwidth]{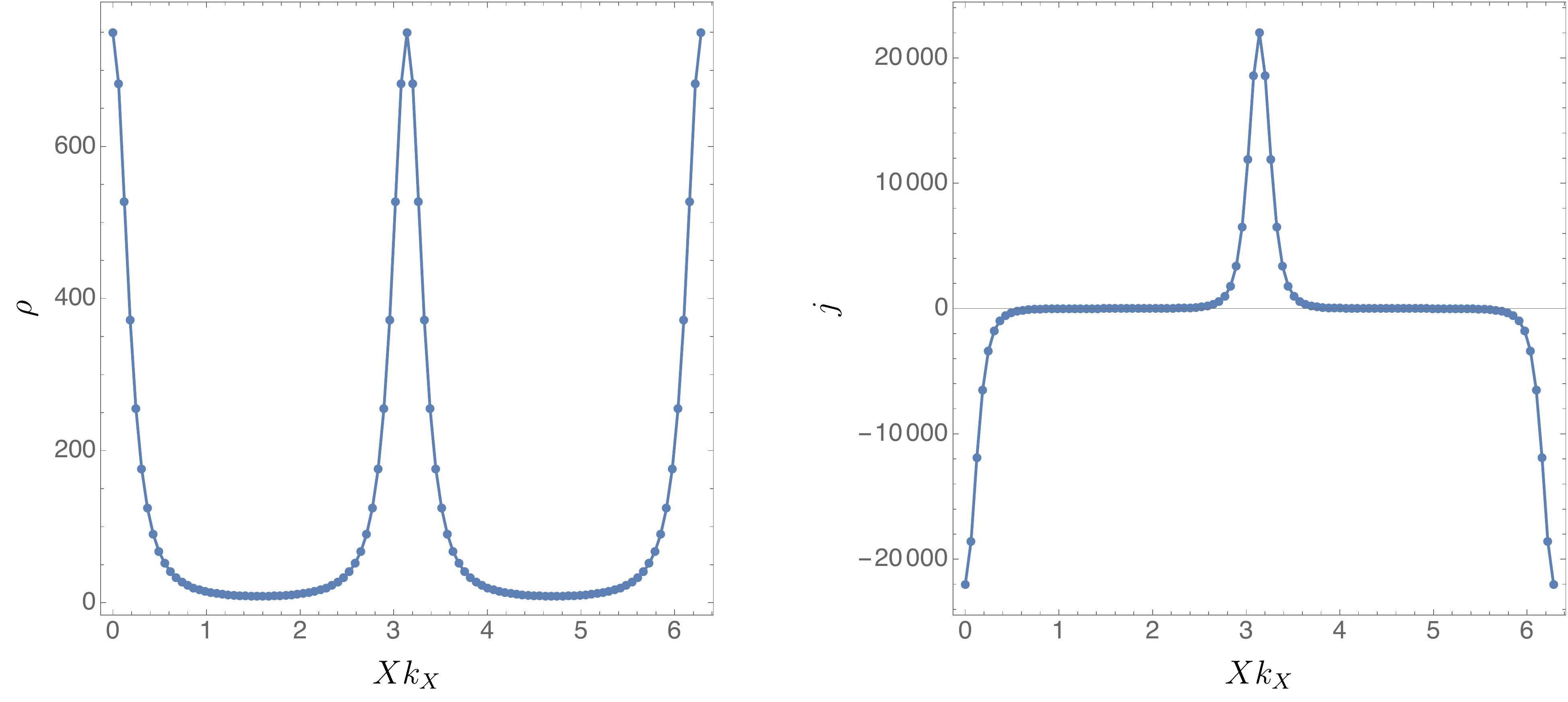}
  \caption{The holographic energy density (left panel) and holographic momentum density (right panel) computed for $a=0.975$  and $T/k_X\approx 1.43$.}
\label{fig:holont1}
\end{figure}

\begin{figure}
\centering
    \includegraphics[width=0.45\textwidth]{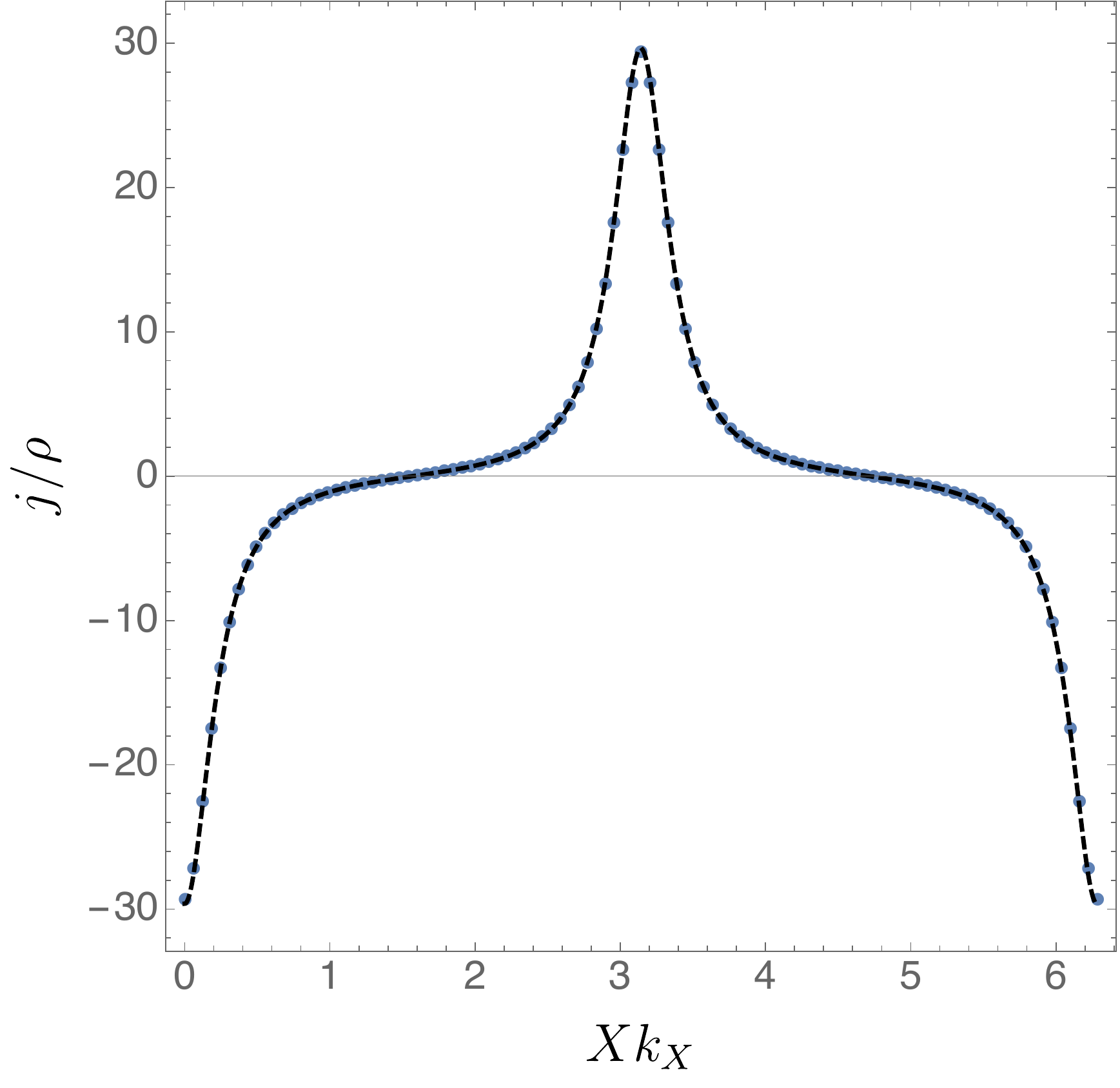}
  \caption{The ratio $j/\rho$ as a function of $X k_X$, computed using $a=0.975$  and $T/k_X\approx 1.43$. The dashed black line represents the fluid calculation and the blue disks our numerical data.}
\label{fig:holont2}
\end{figure}

At high temperatures, the behaviour of $\rho$ and $j$ is dramatically different from the one depicted in Fig.~\ref{fig:holont}. In Fig.~\ref{fig:holont1} we show both $\rho$ and $j$ computed for $T/k_X\approx 1.43$ and $a=0.975$. Although the boundary metric now does not have an ergoregion, if $a$ is increased slightly an ergoregion forms at $k_X X = 0,\pi$.  Note that both the energy and momentum densities develop large features precisely at the location of the would be ergoregion. This is in perfect agreement with the fluid gravity calculation, which indicates a similar feature. 
In Fig.~(\ref{fig:holont2}) we show the analytic curve derived in the fluid approximation (represented as a dashed line) and our numerical data (represented as blue disks). The agreement even at these modest values of $T/k_X$ is very reassuring.

\begin{figure}
\centering
    \includegraphics[width=0.45\textwidth]{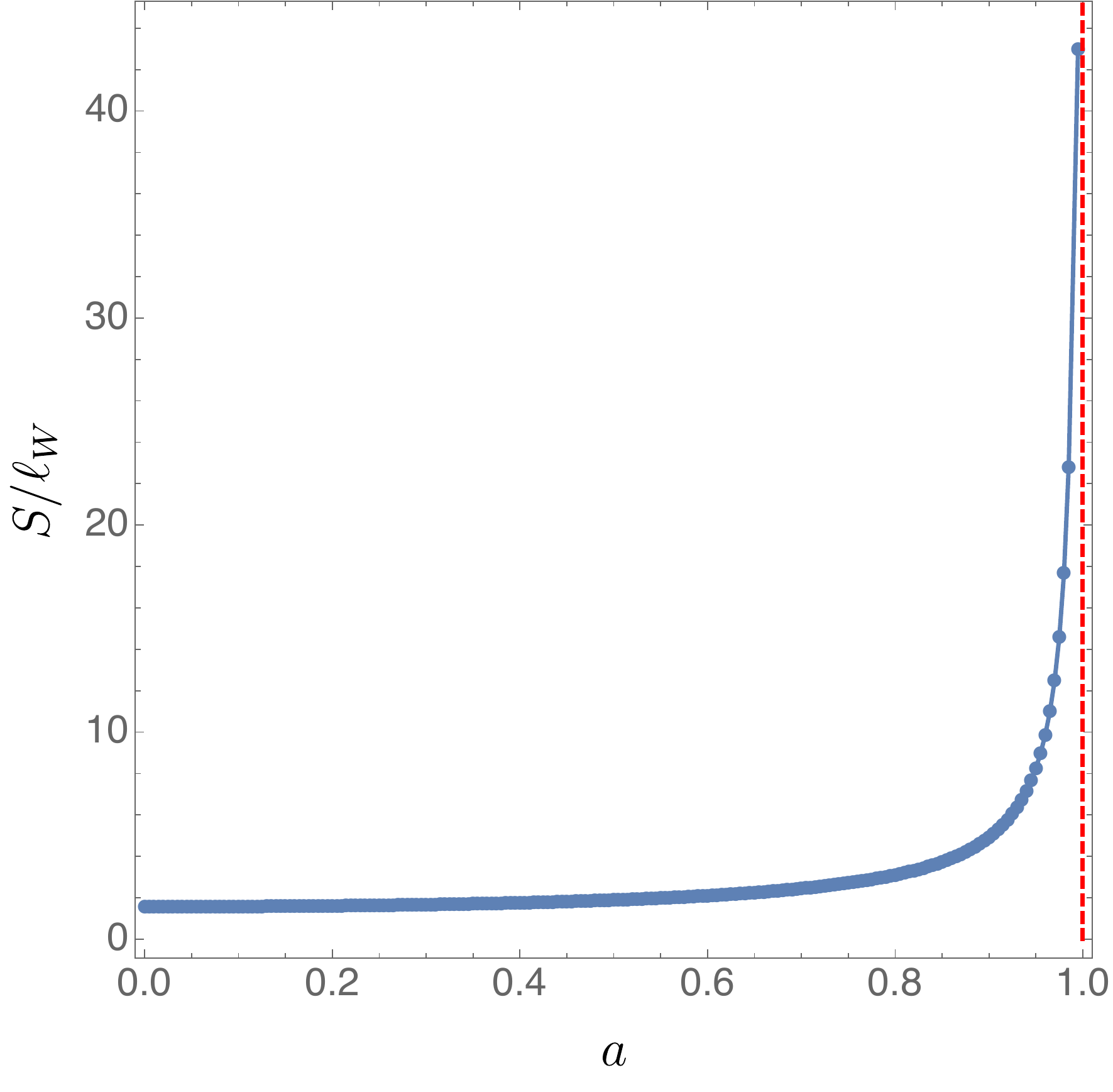}
  \caption{The entropy density $S/\ell_W$ as a function of $a$, computed using $T/k_X\approx0.2387$. The vertical dashed red line marks $a=1$, where a boundary ergoregion forms.
  (This is very close to $a_{\max}$ and our numerics cannot distinguish them.)}
\label{fig:entropy_density}
\end{figure}

If we fix $T/k_X$ and increase $a$, the area of the event horizon increases rapidly as $a\to a_{\max}$. We show this behaviour in Fig.~\ref{fig:entropy_density} where we plot the entropy density $S/\ell_W$ as a function of $a$ and using $T/k_X\approx0.2387$. Other values of $T/k_X$  behave similarly.

\begin{figure}
\centering
    \includegraphics[width=\textwidth]{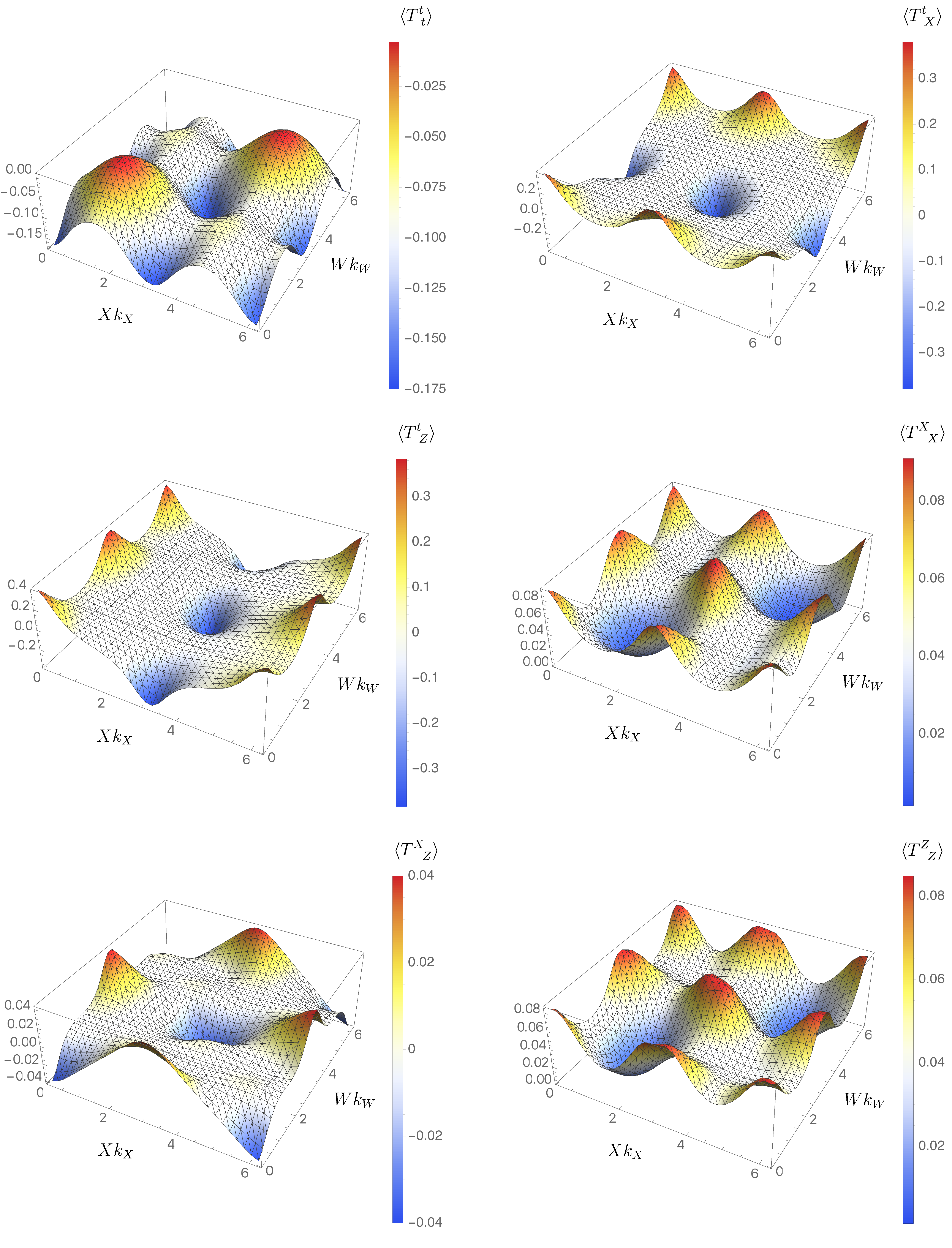}
 \caption{All the components of the holographic stress energy tensor $\langle T^{\mu}_{\;\;\;\nu}\rangle$
 when there is rotation about both circles. These plots are for $T/k_W=0.239$, $k_X =k_W$, and $a=0.6$}
\label{fig:giant_3d}
\end{figure}

Finally, we briefly discuss what happens if we add rotation around both circles and use boundary metric  \eqref{eq:rotcircle2}. 
We will assume equal amplitude for the two rotations: $a = \tilde a$. The calculations are, of course, much more time consuming since we are now solving ten coupled three-dimensional nonlinear partial differential equations. Nevertheless, we reach similar conclusions. Again we find that solutions exist up to a maximum value $a_{\max}$ and that this can be larger than $a_{\ergo}\equiv1/\sqrt{2}$.
The ergoregion now consists of disconnected disks centred at $X k_X = 0,\pi$ and $Wk_W = 0,\pi$ Perhaps the most interesting quantities to plot are now the energy density $\rho$ and the remaining components of the stress energy tensor. These are displayed in Fig.~\ref{fig:giant_3d} for $T/k_W=0.239$, $k_X/k_W=1$, and  $a=0.6$. The behaviour is very similar for other values of $a$ we have studied, except that the extrema get more noticeable as one approaches $a=a_{\max}$. 

\section{Discussion}

We have numerically constructed stationary, asymptotically AdS solutions of Einstein's equation with $\Lambda < 0$ with rotating boundary metrics. If we fix the profile of the differential rotation and increase the overall amplitude $a$, we find a maximum value $a_{\max}$ where the solution becomes singular. This happens both at zero and nonzero temperature, and for compact or noncompact boundaries. We expect that in  the time dependent problem where $a$ is increased from $a <  a_{\max}$  to $a > a_{\max}$  the curvature will grow without bound violating weak cosmic censorship.

However, the boundary metrics all develop ergoregions before reaching $a_{\max}$ and since one can extract energy from an ergoregion, it is natural to ask if there is a positive energy theorem for these boundary conditions.  The following argument suggests that the answer is no. One can clearly place test particles in the ergoregion and boost them so that their energy is arbitrarily negative. We now want to replace the test particle by a small black hole.  There are gluing theorems which ensure that one can add a small black hole  in the ergoregion  to initial data on a constant $t$ surface \cite{Isenberg:2005xp}. An $O(1)$ boost of this black hole will cause it to contribute negatively to the total energy, and should not result any singularities in the initial data. Moving the black hole farther into the asymptotic region increases its negative contribution to the energy without bound.  

Using gauge/gravity duality, there is another argument that the total energy is  unbounded from below. Consider the planar case with $a_{\ergo} < a< a_{\max}$. 
 The boundary metric is an asymptotically flat spacetime with an ergoregion and no horizon. Consider first classical or free quantum fields. Classical fields on such spacetimes are  known \cite{Friedman1978}  to be unstable since one can construct negative energy solutions by exciting fields in the ergoregion. Since stationary solutions must have zero energy and the energy radiated to infinity is always positive, if the energy is negative initially, it will continue to decrease. Free quantum fields on such a spacetime exhibit a similar instability: it has been shown \cite{Ashtekar1975,Kang:1997uw} that there is no Fock vacuum that is time translation invariant. In other words,  there is particle creation in all states. It is also clear that there is no lower bound on the energy for free quantum fields in such a spacetime. This is because excitations localized in the ergoregion can have negative energy, and one can give them arbitrarily large occupation number. 

Even at strong coupling, a CFT on a spacetime with ergoregion and no horizon cannot have a minimum energy state\footnote{We thank D. Marolf for suggesting this argument.}. Start with the ground state in Minkowski space and act with a unitary operator in a finite region $A$. This creates a state with $E>0$ that looks like the vacuum outside $A$. By scale invariance, we can make $A$ as small as we want. Now consider our boundary metric  and pick a small locally flat region inside the ergoregion. As long as $A$ is small enough, we can insert the above state into this geometry. We can then boost it to give it arbitrarily negative energy.

In addition to the instability associated with the ergoregion, there is
another potential instability in the dual field theory if the  scalar curvature is negative over a large enough region.   Conformally invariant scalars in such backgrounds can be unstable. However this is not a problem for the boundary metrics we consider. In the compact case, the scalar curvature of \eqref{eq:rotcircle} or \eqref{eq:rotcircle2} is nonnegative. In the noncompact case, although the scalar curvature of \eqref{eq:bndmetric} can become negative, it is confined to a small area.

We conclude with a comment about another possible class of solutions. In the electromagnetic case, there is a family of static, $T=0$ solutions for any amplitude that describe hovering black holes \cite{Horowitz:2014gva}. These are extremal spherical black holes that hover above the Poincar\'e horizon since the usual attraction to the horizon is balanced by an electrostatic attraction to the boundary. This family of solutions did not play any role in our counterexamples to cosmic censorship since if we only have a Maxwell field, there is no charged matter, and no way to form a charged black hole. It is natural to ask if an analogous hovering black hole could form in the vacuum case and provide a stationary endpoint for any amplitude. It appears the answer is no. We have seen that the singularity arises off the axis so it actually forms a ring. This could not be enclosed by a spherical black hole unless the black hole was quite large and unlikely to be supported by any spin-spin forces.


\vskip 1cm
\centerline{\bf Acknowledgements}
It is a pleasure to thank D.~Marolf and J.~Markevi\v{c}i\={u}t\.e for discussions. GH was supported in part by NSF grant PHY-1504541. JES was supported in part by STFC grants PHY-1504541 and ST/P000681/1.
\vskip .5 cm


\bibliographystyle{JHEP}
\bibliography{all}
  
\end{document}